\documentclass[aps,pra,twocolumn,groupedaddress]{revtex4-2}

\usepackage{booktabs}
\usepackage{amsmath}
\usepackage{graphicx}
\usepackage{verbatim}
\usepackage{subfigure}
\usepackage[english]{babel}
\usepackage{verbatim}
\usepackage{multirow}
\usepackage{xcolor}
\usepackage[utf8]{inputenc}
\usepackage{braket}
\usepackage{bm}
\usepackage{float} 

\newcommand{\ThreeJ}[6]{\left( \begin{matrix}  #1 & #2 & #3 \\ #4 & #5 & #6 \end{matrix} \right)}
\newcommand{\SixJ}[6]{\left \{ \begin{matrix}  #1 & #2 & #3 \\ #4 & #5 & #6 \end{matrix} \right \} }

\begin{document}

\title{The Stark effect in molecular Rydberg states: Calculation of Rydberg-Stark manifolds of H$_2$ and D$_2$ including fine and hyperfine structures}

\author{I. Doran\textsuperscript{1}}

\author{L. Jeckel\textsuperscript{1}}

\author{M. Beyer\textsuperscript{2}}
\author{Ch. Jungen\textsuperscript{3}}

\author{F. Merkt\textsuperscript{1,4,5}} 
\email{frederic.merkt@phys.chem.ethz.ch (F. Merkt)} 

\affiliation{%
  \textsuperscript{1}Institute of Molecular Physical Science, ETH Z\"{u}rich, 8093 Z\"{u}rich, Switzerland%
}

\affiliation{%
  \textsuperscript{2}Department of Physics and Astronomy, LaserLaB, Vrije Universiteit Amsterdam, de Boelelaan 1081, 1081 HV Amsterdam, The Netherlands%
}

\affiliation{%
  \textsuperscript{3}Universit\'{e} Paris-Saclay, Centre Nationale de la Recherche Scientifique, Laboratoire Aim\'{e} Cotton, 91400 Orsay, France
}

\affiliation{%
  \textsuperscript{4}Department of Physics, ETH Z\"urich, Z\"urich, Switzerland}
  
\affiliation{%
  \textsuperscript{5}Quantum Center, ETH Z\"urich, Z\"urich, Switzerland}

\date{\today}

\begin{abstract}
We present a general theoretical treatment and calculations of the fine and hyperfine structures in the spectra of high-$n$ molecular Rydberg states in static uniform electric fields. The treatment combines (i) multichannel quantum-defect theory and long-range polarization models to determine the field-free energies of $n\ell$ Rydberg states of the molecules ($\ell$ is the orbital-angular-momentum quantum number of the Rydberg electron), (ii) a matrix-diagonalization approach to calculate the Stark shifts including their hyperfine structure, and (iii) sequences of angular-momentum frame transformations to predict the line positions and intensities in Stark spectra as they would be observed in single or multiphoton excitation sequences. To clarify how the molecular rotation and the nuclear spins influence the fine and hyperfine structure of molecular Rydberg-Stark spectra, we compare calculated spectra of ortho-D$_2$ with a D$_2^+$ ion core in the rotational ground state ($N^+=0$) for total nuclear spins $I$ of 0 (i.e., without hyperfine structure) and 2 (i.e., with hyperfine structure) with the corresponding spectra of para-H$_2$ with an H$_2^+$ ion core in the first excited rotational state ($N^+=2$) but zero nuclear spin ($I=0$). The calculations show that the hyperfine interaction alone does not significantly modify the Stark effect, but splits each Stark state by almost exactly the hyperfine Fermi-contact splitting of the ion core. In contrast, the effect of the molecular rotation, which is coupled both to the ion-core electron spin by the magnetic spin-rotation interaction and to the Rydberg-electron orbital motion by the core-polarization and charge-quadrupole interactions, induces Stark-state specific splittings that significantly differ from the spin-rotation splitting of the ($N^+=2$) ion core.

\end{abstract}

\maketitle

\section{Introduction}
The Stark effect in atomic and molecular Rydberg states plays a central role in the state-selective detection of Rydberg atoms and molecules by field ionization \cite{stebbings83a,gallagher94a}, in the accurate determination of ionization energies \cite{muellerdethlefs91b,dietrich96a,muellerdethlefs98a, hollenstein01a, hoelsch22a,scheidegger23a,doran24a,scheidegger24a}, in quantum control and manipulation of interatomic interactions \cite{vogt07a,comparat10a,saffman10a,nipper12a,weber17a,stecker20a,jiao22a,mehaignerie25a}, and in electric-field-sensing applications \cite{frey93a,klar94a,osterwalder99a,thiele14a,thiele15a,sedlacek16a,facon16a,zhang24a}. For this reason, it has been extensively studied both experimentally and theoretically.

Rydberg states with a closed-shell ion core such as those of H and the alkali-metal atoms exhibit simple, characteristic spectral structures in the presence of electric fields \cite{zimmerman79a,gallagher94a}. At zero field, Rydberg states of high principal quantum number are characterized by high spectral densities of near degenerate nonpenetrating states with orbital angular momentum quantum number $\ell \ge 4$ and spectrally isolated penetrating states with $\ell < 4$. Whereas the high-$\ell$ Rydberg states exhibit a linear Stark effect at low electric fields already and fan out into characteristic linear Stark manifolds, low-$\ell$ Rydberg states are subject to a quadratic Stark effect at low fields and gradually merge into the linear high-$\ell$ Stark manifolds as the electric field strength increases. If the fine and hyperfine structures are neglected, the magnetic orbital quantum number $m_\ell$ is a good quantum number when the electric field is homogeneous. There are $n-|m_\ell|$ Stark states in each $n, m_{\ell}$-manifold and the states with $|m_\ell|>0$ form pairs of degenerate levels with $m_\ell=\pm | m_\ell |$. The Stark states are labeled by the electric quantum number $k$ \cite{bethe57a}, which takes values ranging from $-(n-|m_\ell|-1)$ to $(n-|m_\ell|-1)$ in steps of 2, see Ref. \cite{zimmerman79a} for the alkali-metal atoms and Fig. \ref{fig:starkmaps}a below for the case of ortho-D$_2$ Rydberg states with zero nuclear-spin and rotational angular momenta. Hyperfine and spin-orbit interactions of the Rydberg electron lead to weak splittings of the levels which very rapidly decrease with increasing $n$ values and are only observable at ultrahigh resolution \cite{scheidegger23a}.

Rydberg atoms with open-shell ion cores have more complex structures because of additional exchange and spin-orbit interactions between the core and Rydberg electrons. The good quantum number is the magnetic quantum number $M_J$ associated with the total electronic angular momentum $\vec{J}$ rather than $m_\ell$, where $\vec{J}=\vec{J}^+ + \vec{j}$ is the sum of the ion-core ($\vec{J}^+$) and Rydberg-electron ($\vec{j}$) total electronic angular momenta. The Stark manifolds typically retain the general structure discussed above, but the Stark states split into multiple magnetic sublevels, see, e.g., Refs. \cite{ernst88a,brevet90a,fielding92a,vrakking97a,merkt98a,shubert11b} for the rare-gas atoms. 

Further complexity is expected to arise when the ion core has a nonzero nuclear spin $\vec{I}$ and its level structure is split by the hyperfine interaction. In this case, which has not yet been studied experimentally, the good quantum number in a homogeneous field is the magnetic quantum number $M_F$ associated with the total angular momentum $\vec{F}=\vec{J}+\vec{I}$. The energy-level pattern in the presence of an electric field then depends on the degree of angular-momentum uncoupling induced by the electric field, in analogy to the uncoupling caused by magnetic fields (Paschen-Back effect). In the limit of full uncoupling, which arises at large field strengths and high $n$ values, the Stark states should form independent manifolds for each hyperfine level of the ion core. The effect of the hyperfine interaction in atomic Rydberg states in electric fields has only been treated theoretically for low-$n$ states in an early study based on perturbation theory and angular-momentum algebra \cite{angel68a}. The tensor formalism introduced in Ref. \cite{angel68a} is still used to calculate static and dynamical polarizabilities and Stark shifts in the ground and low-lying electronic states of atoms \cite{amini03a,bennett99a,simon98a,leeuwen84a} and was used to analyze the Stark effect in $n=40$ and 60 Rydberg states of Cs \cite{herrmann86a}. 

The Stark effect in Rydberg systems has also been extensively studied theoretically in the past decades. Collision-theory approaches based on multichannel quantum-defect theory \cite{sakimoto86a,sakimoto89a,harmin81a,harmin84a,fielding92a,armstrong94a,giannakeas16a} have been applied to treat the Stark effect in the Rydberg states of the rare gas atoms \cite{fielding92a,softley97b,gruetter08a}, molecular hydrogen \cite{fielding91b,fielding94a} and NO \cite{goodgame02a}, however, without consideration of the nuclear spins. More commonly, the Hamiltonian 
\begin{equation}
\hat{H} = \hat{H}_0  + e \mathcal{F} \hat{z}
\label{eq:htotstark_intro}
\end{equation}
is expressed in matrix form, where $\hat{H}_0$ describes the atom or molecule in the absence of the electric field and the Stark effect is included through the additional potential-energy term $ e \mathcal{F} \hat{z}$ perceived by the Rydberg electron in the external electric field $\vec{F}=(0,0, \mathcal{F} )^\dagger $ \cite{littman76a,zimmerman79a,stebbings83a,gallagher94a}. The Stark spectrum is obtained by determining the eigenvalues of the matrix under conditions where a sufficiently large range of $n$ values are included to reach convergence. In both approaches, the effect of the external electric field on the ion-core is neglected, which is an excellent approximation at low fields and for ion cores in low-lying electronic states. Such approaches have been used to quantitatively account for the Stark effect in atomic systems with full-shell (see, e.g., \cite{tauschinsky13a,grimmel15a,sibalic17a,peper19b,duspayev24a} for alkali-metal atoms) and open-shell ion cores (see, e.g., \cite{ernst88a,brevet90a,vrakking97a,merkt98a} for the rare-gas atoms).

The same approaches can in principle also be applied to the Stark effect in \textit{molecular} Rydberg states, but progress has been slower. The level structures of molecular ions consist of a multitude of rovibrational levels, which are further split by the spin-orbit and hyperfine interactions. Rydberg series of different $\ell$ values and converging on the different rovibrational and hyperfine levels of the ion core interact, leading to dense and highly perturbed level structures already at zero field \cite{jungen69a,herzberg72a,freund83a,ginter80a,ginter84a,ginter88a,huber90a,jungen96a,jungen11a,vervloet88a,seiler03a,deller20b,langford98a,raptis99a,field05a,kay11a,raptis01a,bacon00a}. These series are further coupled by the external electric field. Moreover, the angular-momentum coupling must be considered in both the space-fixed and molecule-fixed coordinate systems, which results in a high level of complexity and a large number of limiting cases of angular-momentum (un)coupling. The collision and matrix-diagonalization approaches outlined above have both been applied to interpret molecular Rydberg-Stark spectra in molecular hydrogen \cite{fielding91b,fielding94a,seiler11b,hoelsch22a}, NO \cite{vrakking96a, warntjes99a,goodgame02a, clarson08a, deller20a, rayment21a,munkes24a}, Na$_2$ \cite{bordas87a}, N$_2$ \cite{softley97b}, H$_3$ \cite{bordas92a, bordas93a, menendez06a}, BaF \cite{zhou15a} and CaF \cite{petrovic09a}, however neglecting the effects of electron and nuclear spins. This neglect was entirely justified because no experimental data had been obtained at a spectral resolution sufficient to resolve the hyperfine structure of Rydberg-Stark spectra.

We present here an extension of the matrix-diagonalization approach to the characterization of the Stark effect in molecular Rydberg states under conditions where the hyperfine structure is resolved. Our motivations for this extension are twofold. First, we would like to reach a global understanding of penetrating and nonpenetrating molecular Rydberg states and how they are affected by the combined effects of fine and hyperfine interactions and electric fields. Measuring Rydberg spectra in the presence of electric fields provides access to high-$\ell$ states which are otherwise very difficult to study under field-free conditions because of restrictions imposed by optical selection rules. The electric field decouples the rotational, orbital, electron spin and nuclear-spin angular momenta and is expected to modify the angular-momentum-coupling hierarchy, leading to rich spectral patterns and to information not obtainable under field-free conditions. 

Secondly, we would like to use the Stark effect as a means to determine precise values of ionization energies and fine- and hyperfine-structure intervals in molecular ions. We have determined the ionization energies of para-H$_2$ corresponding to the formation of H$_2^+$ in the $v^+=0, N^+=0$ and  $v^+=1, N^+=0$ levels with high accuracy from the analysis of the corresponding Rydberg-Stark spectra of H$_2$ at principal quantum numbers in the range between 45 and 70 \cite{hoelsch22a,doran24a}. These levels do not have rotational angular momentum and the nuclear spin of para-H$_2$ is 0. Consequently, the level structure is not affected by spin-rotation and hyperfine splittings. We are currently extending these measurements to the $v^+=1, N^+=2$ and the $v^+=1, N^+=0$ rovibrational levels of H$_2^+$ and D$_2^+$, respectively \cite{iodo_tbp}. The interpretation of the spectra requires understanding the Stark effect including the fine and hyperfine structures caused by spin-rotation and hyperfine interactions and an extension of the theoretical framework to rigorously include all relevant interactions. The description of this extension is the main purpose of the present article. 

\section{Method}
\label{sec:general}
The energy levels of molecular Rydberg states in the presence of a homogeneous electric field are calculated expressing the total effective Hamiltonian operator in Eq. \eqref{eq:htotstark_intro} in matrix form in a basis set which labels each Rydberg state by a unique set of quantum numbers, appropriate for a given angular-momentum coupling scheme. The quantum numbers of the angular momenta used in this work are listed in Table \ref{tab:qnr}. Relevant considerations regarding the Rydberg states studied here are that i) the molecular ion core is in its X$^+$ $^2 \Sigma_g ^+$ electronic ground state with $\Lambda^+=0$ (such that the projection of the total orbital angular momentum onto the internuclear axis corresponds to $\Lambda = \Lambda^+ + \lambda = \lambda$); ii) the molecular ion core is homonuclear (H$_2^+$ or D$_2^+$), with two distinct nuclear-spin isomers: ortho-D$_2^+$ ($I$ = 0,2, even $N^+$) and para-D$_2^+$ ($I$=1, odd $N^+$), and para-H$_2^+$ ($I$=0, even $N^+$) and ortho-H$_2^+$ ($I=1$, odd $N^+$). 

\begin{figure*}
    \centering
     {\includegraphics[trim=3cm 17cm 0cm 7cm, clip=true, width=\linewidth]{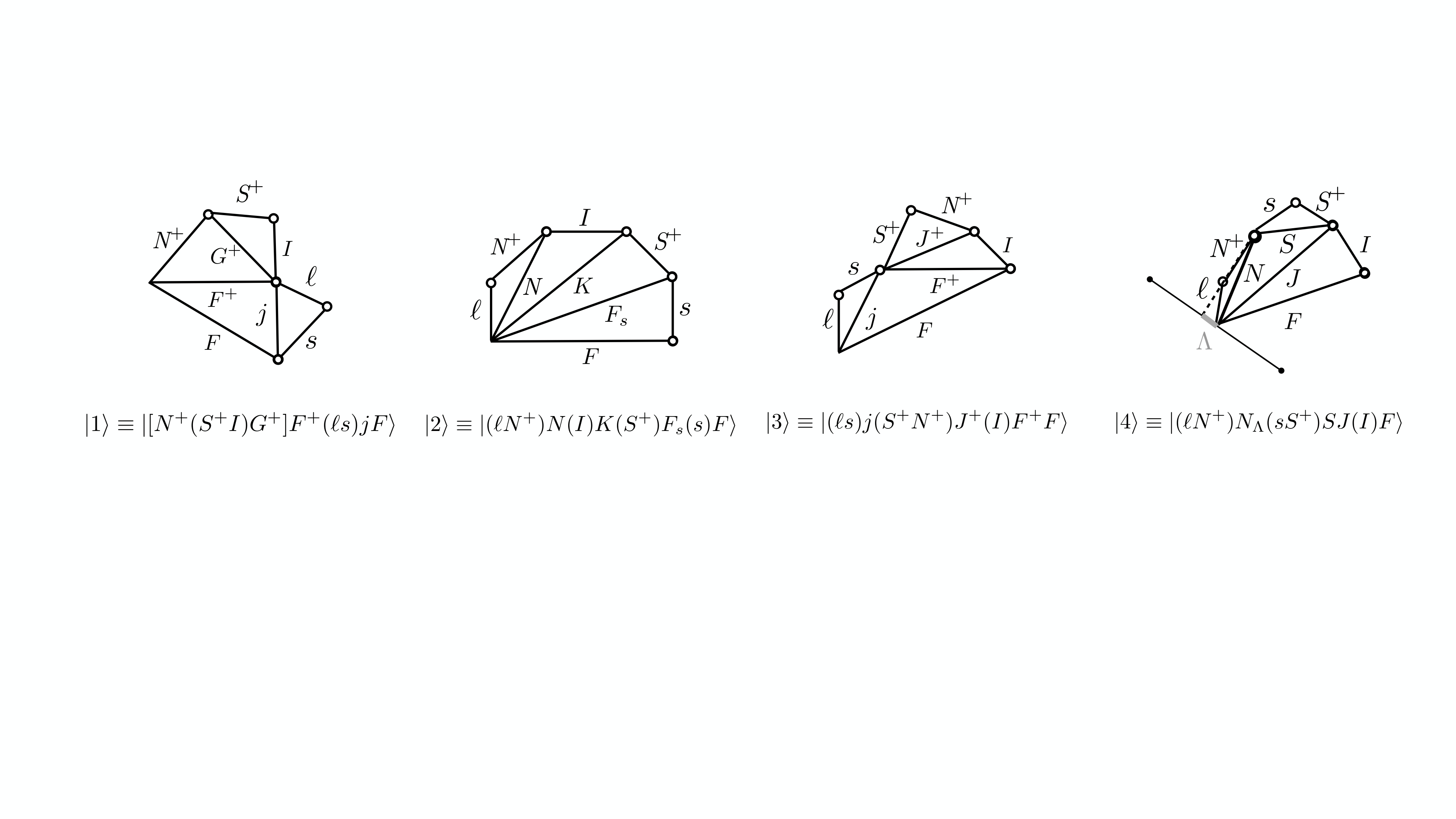}}
      \caption{Angular-momentum coupling diagrams showing the basis sets of matrices used in the calculations of the: MQDT energies ($|1 \rangle$, Section \ref{subsec:mqdt}), long-range interaction model ($|2 \rangle$, Section \ref{subsec:longrange}), Stark effect ($|3 \rangle$, Section \ref{sec:stark}), intensities ($|4 \rangle$, Section \ref{sec:intensity_model}).}
      \label{fig:basis_sets}
\end{figure*}

To investigate the effects of the nuclear spin and of the rotational angular momentum separately, we focus on the $N^+=0$ Rydberg states of ortho-D$_2$ ($I=2$) and on the $N^+=2$ Rydberg states of para-H$_2$ ($I=0$). The calculation of Rydberg-Stark spectra requires the calculation of the field-free energies (Section \ref{sec:fieldfree}), of the electric-field-induced couplings (Section \ref{sec:stark}), and of the line positions and intensities of the Stark spectra corresponding to a chosen multiphoton excitation scheme (Section \ref{sec:intensity_model}). The approximations made in the calculations and the resulting uncertainties are presented in Section \ref{sec:errors}. The different steps of the calculations are performed using different basis sets corresponding to specific angular-momentum-coupling hierarchies (see Fig. \ref{fig:basis_sets}), for the reasons given in the corresponding sections. In the designation of these basis sets, the symbols in parentheses indicate the 
angular momenta that are coupled to give the angular momentum on the right of the parentheses. For instance $|2\rangle$ = $\ket{ (\ell N^+)N(I)K(S^+)F_s(s)F}$ implies that $\vec{\ell}$ and $\vec{N}^+$ are coupled to yield $\vec{N}$, which is then coupled to $\vec{I}$ to yield $\vec{K}$, and so on. The frame transformations used to connect the different basis sets in Fig. \ref{fig:basis_sets} are given in Section \ref{sec:frametrafos} and in Appendix A. They extend transformations derived by Jungen and Raseev \cite{jungen98a} to the treatment of the hyperfine structure.

\begin{table}[h]
\centering
\caption{List of angular-momentum quantum numbers used in this work. Intermediate quantum numbers (e.g., $N, K, F_s$ for $|2\rangle$) are defined in Fig. \ref{fig:basis_sets}.}
\label{tab:qnr}
\begin{tabular}{lc} 
\toprule
Symbol & Physical quantity \\
\midrule
$\ell$ & Orbital angular momentum of Rydberg electron \\
$\lambda$ & Projection of $\ell$ onto internuclear axis \\
$s$ & Spin of Rydberg electron \\ 
$N^+$ & Rotational angular momentum of ion core \\ 
$I$ & Nuclear spin \\
$S^+$ & Spin of ion-core electron \\
$\Lambda^+ = 0$ & Projection of orbital angular momentum \\
& of ion-core electron onto internuclear axis \\
$\Lambda$ & Projection of total orbital angular \\
& momentum onto internuclear axis \\
$F$ & Total angular momentum \\
$M_F$ & Projection of $F$ onto the $z$-axis of \\
& laboratory-fixed frame\\
\bottomrule
\end{tabular} 
\end{table}

\section{Calculation of Rydberg-state energies at zero field}
\label{sec:fieldfree}
\subsection{MQDT calculations  for low-$\ell$ Rydberg states}
\label{subsec:mqdt}

A complete and accurate theoretical description of molecular Rydberg states under field-free conditions is provided by multichannel-quantum-defect-theory (MQDT) \cite{seaton66a, fano70a, seaton83a, greene85a, jungen96a, jungen11a}. MQDT is an extended scattering theory which describes Rydberg states as bound levels in collisions between the Rydberg electron and the ion core. Frame transformations \cite{fano70a, jungen11a} are used to connect the short-range coupling regime (corresponding to the \textit{ab-initio} calculations of potential energy curves of low-$n$ Rydberg states, for which the Rydberg electron wavefunction overlaps with the ion core) to the long-range coupling regime, appropriate for a global description of high-$n$ Rydberg states, for which the electron-ion-core interaction can be described by the long-range electrostatic interactions. MQDT describes the ionization channels and their interactions. An ionization channel consists of the bound and continuum states associated with a given ion-core quantum state and an outer electron of well-defined $\ell$ value. The use of MQDT to determine the field-free energies of s, p, d and f Rydberg states needed to obtain $\hat{H}_0$ in molecular hydrogen has been described elsewhere \cite{hoelsch22a,osterwalder04a,sprecher14b}. In brief, short-range quantum-defect parameters are used, which describe the dependence of the quantum defects on the collision energy and the internuclear separation $R$ for each set of $S=0-1, \ell=0-3$ and $\lambda=0-\ell$ values. These short-range quantum defects are assumed to be independent of nuclear spins \cite{woerner03a, osterwalder04a}. In the present work, MQDT calculations including electron and nuclear spins were performed for $\ell=0-3$ states using the programs developed by Christian Jungen and coworkers, as described in Refs. \cite{osterwalder04a, sprecher14b}. The MQDT calculations were carried out in energy regions corresponding to the range of principal quantum number $n$ suited for precision experiments ($\approx$ 30 to 50) \cite{iodo_tbp}, including all rotational channels and vibrational channels up to $v^+=8$. For each calculated energy level, the corresponding eigenvector is expressed as a linear combination of contributing \textit{dissociation channels} $\ket{v^+} \ket{\left[ N^+ (S^+ I)G^+ \right] F^+ (\ell s) j F}$, corresponding to angular-momentum basis set $\ket 1$ in Fig. \ref{fig:basis_sets}. In these channels, the angular momenta of the ion core ($\vec{N}^+, \vec{S}^+$ and $\vec{I}$) and the Rydberg electron ($\vec{\ell}$ and $\vec{s}$) are coupled separately, to give $\vec{F}^+$ and $\vec{j}$, respectively. The relative contributions of different $v^+, N^+$ rovibrational ionization channels characterize the level of channel mixing, related to nonadiabatic rovibrational interactions, affecting the Rydberg states of interest. In this work, the series used to illustrate the calculation procedure are primarily associated with $v^+=1, N^+=0$ for ortho-D$_2$ and $v^+=1, N^+=2$ for para-H$_2$ (see also Section \ref{sec:general}). 

In a first step, the MQDT calculations were used to identify the regions where rovibrational channel interactions are not significant. In these regions, the basis set for the calculations can be restricted to specific $N^+$ and $v^+$ values, i.e., $N^+=0, v^+=1$ for ortho-D$_2$ and $N^+=2, v^+=1$ for para-H$_2$, which simplifies the subsequent calculations of the Stark effect without loss of generality. The matrix elements for $\hat{H}_0$ are derived from the MQDT-given eigenvalues and their corresponding eigenvectors in the $\ket{\left[ N^+ (S^+ I)G^+ \right] F^+ (\ell s) j F}$ basis (basis $\ket{1}$ in Fig. \ref{fig:basis_sets}). In this basis, the core-total-spin quantum number $G^+$ is a good quantum number.
A frame transformation is performed to the $\ket {(\ell s)j \left[ (S^+ N^+)J^+(I) \right] F^+F}$ basis ($\ket{3}$ in Fig. \ref{fig:basis_sets}), which is then used to evaluate the field-induced interactions (Section \ref{sec:stark}). The explicit expression for this frame transformation is given by Eqs. (IIA-C) in the Appendix. The quantum number $M_F$ can be omitted in the notation because all $M_F$ components of a given state are degenerate in the absence of external fields.

\begin{figure*}
    \centering
     {\includegraphics[trim=0cm 0cm 0.5cm 0cm, clip=true, width=\linewidth]{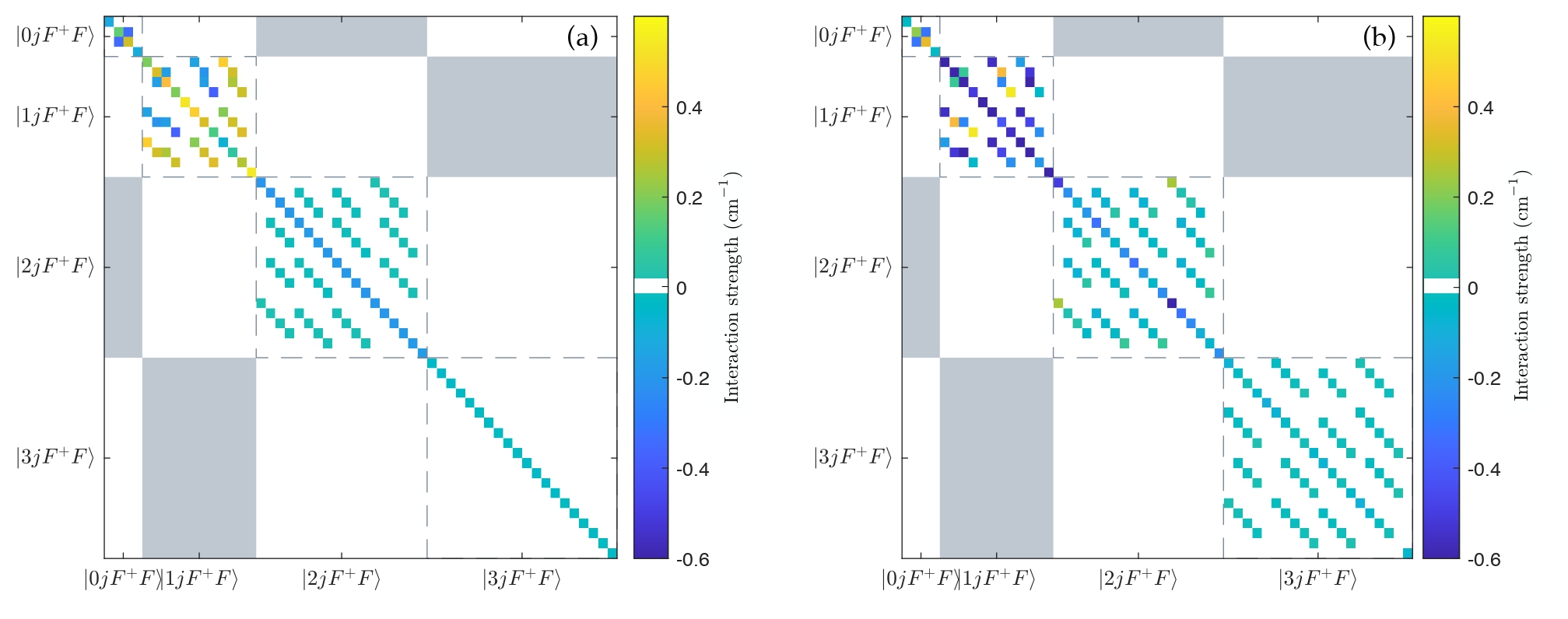}}
      \caption{$\ell=0-3, n = 34, v^+=1$ zero-field matrices for (a) ortho-D$_2$, $I=2, N^+=0$ and (b) para-H$_2$, $I=0, N^+=2$ obtained from MQDT calculations including spins. The different entries within each diagonal block correspond to the possible values of $F$ resulting from the addition of $\vec{j}$ and $\vec{F}^+$, e.g., in the case of $\ell=0$, $F=1$ for $j=1/2$ and $F^+=3/2$, $F=2$ for $j=1/2$ and $F^+=3/2$ and $F^+=5/2$, and $F=3$ for $j=1/2$ and $F^+=5/2$. The diagonal elements are given with respect to $- R_{\textrm{D}_2 (\textrm{H}_2)} / {n^2}$. The blocks are labeled by $\ket{\ell jF^+F}$.}
    \label{fig:matrices_Hzero_oD2_pH2}
\end{figure*}

To illustrate typical results of the MQDT calculations, Fig. \ref{fig:matrices_Hzero_oD2_pH2}a and b show the entries of $\hat{H}_{0, \ell \leq 3}$ corresponding to $n=34$ and $\ell = 0-3$ after the frame transformation to the $\ket {(\ell s)j \left[ (S^+ N^+)J^+(I) \right] F^+F}$ basis for ortho-D$_2$ ($I=2, v^+=1, N^+=0$), and para-H$_2$ ($I=0, v^+=1, N^+=2$), respectively. The rows and columns are labeled as $\ket {\ell j F^+ F}$, and for conciseness only the value of $\ell$ is explicitly given (complete labels for all states are given in Figs. S1 and S2 of the Supplemental Material). The diagonal elements are given with respect to the respective Bohr energies $- R_{\textrm{D}_2 (\textrm{H}_2)} / {n^2}$. The matrices are to an excellent approximation block-diagonal in $\ell$. The strengths of the s-d and p-f electronic couplings \cite{ross94a, ross94b, ross94c, osterwalder04a} (indicated as light grey rectangles) are negligible on this scale. The dominant couplings are between $\ket {\ell j F^+ F}$ states of the same total quantum number $F$, as indicated by the off-diagonal elements within each block-diagonal submatrix of $\ell=0-2$. These couplings are caused by the exchange interaction and the electrostatic coupling of $\vec{\ell}$ with $\vec{N}^+$ (for $N^+ > 0$) in the short-range region, as well as by magnetic interactions between the Rydberg electron and the ion core. In ortho-D$_2$, the  34f block is diagonal in the $\ket {\ell j F^+ F}$ basis because the exchange interaction is negligible and, for $N^+=0$, there are no other significant couplings between the angular momenta of the ion core and the Rydberg electron. In para-H$_2$ ($N^+=2$) on the other hand, the electrostatic coupling between $\vec{\ell}$ and $\vec{N}^+$ ($\vec{N} = \vec{\ell} + \vec{N}^+$) results in off-diagonal elements in the $\ket {\ell j F^+ F}$ representation of the f states.
We restrict the use of MQDT calculations to penetrating $\ell \leq 3$ Rydberg states, for which quantum defects are available either from ab-initio calculations \cite{silkowski21a, silkowski22a,silkowski24a} or from extensive fits to experimental data \cite{osterwalder04a, sprecher14b}. For $\ell \geq 4$ Rydberg states, we calculate the zero-field energies using a long-range interaction model, as discussed in Section \ref{subsec:longrange}.

\subsection{Long-range interaction model for high-$\ell$ Rydberg states}
\label{subsec:longrange}

With increasing $\ell$, the Rydberg electron penetrates less and less into the core region and therefore the effects of channel interactions are reduced. In the following treatment, we neglect channel interactions altogether for $\ell \geq 4$, after verifying in MQDT calculations for $\ell=0-3$ states that the energy regions of interest are not significantly perturbed.

The energies of core-nonpenetrating ($\ell \geq 4$) Rydberg states have been previously calculated using a long-range model which includes i) the charge-quadrupole interaction of the Rydberg electron with the quadrupole moment of the electric-charge distribution in the ion core, as well as ii) the polarization of the ion core by the Rydberg electron (charge--induced-dipole interaction), as introduced by Jungen and Miescher in their study of f Rydberg states of NO \cite{jungen69a}. Higher-order effects arising from the octupole and hexadecapole moments and higher-order polarizability terms, as well as nonadiabatic effects, can also be taken into account \cite{sturrus88a, jungen89a}. Numerous measurements of Rydberg states have served as experimental validation of this model for Rydberg states of molecular hydrogen \cite{herzberg82a, sturrus85a, knight85c, eyler86a, basterrechea93a, uy00a}. Here, we implement a spin-independent long-range interaction model in Hund's angular-momentum coupling case d) $\ket{n v^+}\ket{ (\ell N^+)N }$ \cite{eyler83a}, to which we add the effects of magnetic interactions \cite{sturrus86a, palfrey84a}. The basis set chosen for this purpose is $ \ket{nv^+} \ket{(\ell N^+)N(I)K(S^+)F_s(s)F}$ ($ \ket 2$ in Fig. \ref{fig:basis_sets}), and is built starting from the spin-independent $\ket{nv^+} \ket{ (\ell N^+)N }$ basis functions, to which the nuclear and electron spins $I, S^+$ and $s$ are added consecutively. This choice of basis set is convenient because the diagonal elements correspond to the spin-independent energies of the $\ket{nv^+} \ket{ (\ell N^+)N}$ states, as derived in Ref. \cite{eyler83a} (see also Ref. \cite{jungen69a}).
The overall Hamiltonian operator for the long-range interaction including spins used to describe the $\ell \geq 4$ states is then
\begin{align}
   \hat{H}_{0, \ell \geq 4} = \hat{H}_\textrm{Coulomb} & + \hat{H}_\textrm{lr} + \hat{H}_\textrm{hfs} + \hat{H}_\textrm{spin}^{\textrm{BB}} \notag\\
     \begin{aligned}[t]
        & + \hat{H}_\textrm{exchange}, \\
     \end {aligned} 
     \label{eq:H_LR_tot} 
\end{align}
where the first two terms represent the spin-independent contribution, followed by the magnetic hyperfine ($\hat{H}_{\textrm{hfs}}$), the spin-orbit and spin-spin interactions in the Breit-Bethe approximation ($\hat{H}_\textrm{spin}^{\textrm{BB}}$), and the exchange interaction ($\hat{H}_\textrm{exchange}$). The calculation of these contributions is discussed in detail below.

Neglecting magnetic and exchange interactions, the interaction between the Rydberg electron and the ion core can be written as a multipole expansion and expressed in terms of Legendre polynomials as
\begin{equation}
\hat{H}_\textrm{multipole} = - \frac{1}{4 \pi \epsilon_0} \frac{e Q}{|\bm{r-R_\textrm{c}}|} =   \frac{1}{4 \pi \epsilon_0} \sum_{k} -e Q \frac{R_\textrm{c}^k} {r^{k+1}} P_k (\textrm{cos} \theta),
\label{eq:H_multipole}
\end{equation}
where \textbf{r} and $\bm{R_\textrm{c}}$ ($-e$ and $Q$) are the positions (electric charges) of the Rydberg electron and of the ion-core center of charge. The first term in this expansion represents the charge-charge (Coulomb) interaction 
\begin{equation*}
\hat{H}_\textrm{Coulomb} = - \frac{1}{4 \pi \epsilon_0} \frac{eQ}{r}
\end{equation*}
and results in (diagonal) matrix elements
\begin{widetext}
\begin{equation*}
\bra{nv^+} \braket{(\ell N^+)N(I)K(S^+)F_s(s)FM_F|\hat{H}_\textrm{Coulomb}|(\ell' N^{+}{'})N'(I')K'(S^{+}{'})F_s'(s')F'M_F'} \ket{n'v^{+}{'}}
\end{equation*}
\begin{equation}
=\delta_{nn'} \delta_{v^+ v^{+}{'}} \delta_{\ell \ell'} \delta_{N^+ N^{+}{'}}  \delta_{N N'} \delta_{II'} \delta_{KK'} \delta_{S^+ S^{+}{'}} \delta_{F_s F_s'} \delta_{ss'} \delta_{FF'} \delta_{M_F M_F'} \left[ E_I (v^+,N^+) - \frac{R_{M}}{n^2} \right],
\label{eq:HCoulomb4}
\end{equation}
\end{widetext}
where $R_M = R_\infty \frac{\mu}{m_e}$ is the mass-reduced Rydberg constant ($\mu = \frac{m_eM_\textrm{c}}{m_e + M_\textrm{c}}$). 
In the case of the homonuclear diatomic molecules H$_2$ and D$_2$, the next term in the expansion is the charge-quadrupole interaction, which can be expressed as \cite{jungen69a,eyler83a} 
\begin{equation}
\hat{H}_\textrm{quadrupole} = - \frac{1}{4 \pi \epsilon_0} \frac{eQ(R)}{r^3} P_2(\textrm{cos} \theta),
\label{eq:H_quadrupole}
\end{equation}
where $Q(R) = QR_c^2$ is the (static) quadrupole moment and $\theta$ the polar angle of the Rydberg electron in the molecular frame.

The electric field of the Rydberg electron polarizes the ion core and leads to additional interactions between the Rydberg electron and the resulting induced multipoles in the ion core. The leading contribution to the polarization Hamiltonian is the charge--induced-dipole interaction. It takes the form
\begin{equation}
\hat{H}_\textrm{polarization} = - \frac{1}{4 \pi \epsilon_0} \frac{\alpha e^2}{2r^4} - \frac{1}{4 \pi \epsilon_0} \frac{\gamma e^2}{3r^4} P_2(\textrm{cos} \theta),
\label{eq:H_longrange}
\end{equation}
where $\alpha = \frac{2 \alpha_\perp + \alpha_\parallel}{3}$ and $\gamma = \alpha_\parallel - \alpha_\perp$ are the isotropic and anisotropic polarizabilities of the ion core, expressed as a function of the elements of the polarizability tensor 

\begin{equation*}
\alpha=
\begin{pmatrix}
\alpha_\perp & 0 & 0\\
0 & \alpha_\perp & 0\\
0 & 0 & \alpha_\parallel
\end{pmatrix}
\end{equation*}
in the molecule-fixed coordinate system.
To first order, the spin-independent long-range interaction is $\hat{H}_\textrm{lr} =  \hat{H}_\textrm{quadrupole} + \hat{H}_\textrm{polarization}$ and its diagonal matrix elements are given by \cite{eyler83a} (see also \cite{jungen69a}):
\begin{widetext}
\begin{equation*}
\bra{nv^+} \braket{(\ell N^+)N(I)K(S^+)F_s(s)FM_F|\hat{H}_\textrm{lr}|(\ell' N^{+}{'})N'(I')K'(S^{+}{'})F_s'(s')F'M_F'} \ket{n'v^{+}{'}}
\end{equation*}
\begin{equation*}
=\delta_{nn'} \delta_{v^+ v^{+}{'}} \delta_{\ell \ell'} \delta_{N^+ N^{+}{'}}  \delta_{N N'} \delta_{II'} \delta_{KK'} \delta_{S^+ S^{+}{'}} \delta_{F_s F_s'} \delta_{ss'} \delta_{FF'} \delta_{M_F M_F'} 
\end{equation*}
\begin{equation*}
\times \left[ -  \frac{1}{4 \pi \epsilon_0} \frac{e}{a_0^3} \braket{ v^+ N^+ | Q(R) | v^+ N^+} \frac{1}{n^3(\ell+1)(\ell+\frac{1}{2})\ell} \ \frac{3Y(Y-1)-4N^+(N^++1)\ell (\ell+1)}{2(2\ell-1)(2N^+-1)(2\ell+3)(2N^++3)} \right.
\end{equation*}

\begin{equation*}
- \frac{1}{4 \pi \epsilon_0} \frac{1}{2} \frac{e^2}{a_0^4} \braket{ v^+ N^+ | \alpha(R) | v^+ N^+} \frac{\frac{1}{2}[3n^2-\ell(\ell+1)]}{n^5(\ell+\frac{3}{2})(\ell+1)(\ell+\frac{1}{2})\ell(\ell-\frac{1}{2})}
\end{equation*}

\begin{equation}
\left. -  \frac{1}{4 \pi \epsilon_0} \frac{1}{3} \frac{e^2}{a_0^4} \braket{ v^+ N^+ | \gamma(R) | v^+ N^+} \frac{\frac{1}{2}[3n^2-\ell(\ell+1)]}{n^5(\ell+\frac{3}{2})(\ell+1)(\ell+\frac{1}{2})\ell(\ell-\frac{1}{2})}  \ \frac{3Y(Y-1)-4N^+(N^++1)\ell (\ell+1)}{2(2\ell-1)(2N^+-1)(2\ell+3)(2N^++3)} \right],
\label{eq:Hpol4}
\end{equation}
\end{widetext}
where $Y = N^+(N^++1) + \ell(\ell+1) - N(N+1)$ and $\braket{ v^+ N^+ | Q(R) | v^+ N^+} $, $\braket{ v^+ N^+ | \alpha(R) | v^+ N^+} $, and $\braket{ v^+ N^+ | \gamma(R) | v^+ N^+}$ are rovibrationally averaged values (over the internuclear distance $R$) taken from \textit{ab-initio} calculations \cite{schiller14a,karr25p}. Nonzero off-diagonal elements of $\hat{H}_\textrm{lr}$ with $\Delta N^+ = \pm 2$ and $\Delta \ell = \pm 2$ result from the quadrupole and charge--induced-dipole interactions. However, these elements are much smaller and their effects are negligible in the $n$ regions of interest \cite{iodo_tbp} (see Subsection \ref{subsec:mqdt}).

For the homonuclear molecular hydrogen ions, hyperfine-interaction terms involving the Rydberg electron can be safely neglected in high-$n$ Rydberg states \cite{sun89a, woerner03a, osterwalder04a} and the effective hyperfine Hamiltonian is \cite{korobov06c}
\begin{widetext}
\begin{equation*}
\hat{H}_\textrm{hfs} = b_F (\bm{I} \cdot \bm{S^+}) + c_e(\bm{N^+} \cdot \bm{S^+} )+ c_I(\bm{N^+} \cdot \bm{I}) + \frac{d_1}{(2N^+-1)(2N^++3)} \left[\frac{2}{3} (\bm{N}^{+})^{2} (\bm{I} \cdot \bm{S^+}) - (\bm{N^+} \cdot \bm{I})(\bm{N^+} \cdot \bm{S^+})  \right. 
\end{equation*}
\begin{equation}
\left. - (\bm{N^+} \cdot \bm{S^+})(\bm{N^+} \cdot \bm{I}) \right] + \frac{d_2}{(2N^+-1)(2N^++3)} \left[\frac{1}{3} (\bm{N}^{+})^{2} \bm{I}^2 - \frac{1}{2}(\bm{N^+} \cdot \bm{I}) - (\bm{N^+} \cdot \bm{I})^2 \right],
\label{eq:Hhfs_exp2}
\end{equation}
\end{widetext}
where $b_F, c_e, c_I, d_1$ and $d_2$ are effective $v^+,N^+$-dependent state-specific coefficients. For the calculations presented below, we used the calculated values $b_F = 139.837$ MHz for D$_2^+ (v^+=1, N^+=0)$ \cite{danev21a} and $c_e = 39.5716$ MHz for H$_2^+ (v^+=1, N^+=2)$ \cite{korobov06c}. In the states selected for this investigation, the ion core has either a nonzero nuclear spin ($I=2, N^+=0$ for ortho-D$_2^+$) or a nonzero rotational angular momentum ($I=0, N^+=2$ for para-H$_2^+$). Consequently, there are no terms involving $\bm{N^+}\cdot \bm{I}$, and only the first two terms in Eq. \eqref{eq:Hhfs_exp2} are relevant. The matrix elements of the spin-rotation coupling term $c_e(\bm{N^+} \cdot \bm{S^+} )$ can be evaluated as
\begin{widetext}
\begin{equation*}
c_e \braket{(\ell N^+)N(I)K(S^+)F_s(s)FM_F| \bm{T^1}(N^+) \cdot \bm{T^1}(S^+)|(\ell ' N^{+}{'})N'(I')K'(S^{+}){'}F_s'(s')F'M_F'} 
\end{equation*}
\begin{equation*}
=c_e  \delta_{M_F M_F'}\delta_{FF'} \delta_{F_sF_s'} \delta_{I I'} \delta_{\ell \ell'} \delta_{N^+ N^{+}{'}} \delta_{S^+ S^{+}{'}} (-1)^{K' + S^+ +F_s  + I + K +  \ell + N^{+}{'}} \SixJ{K}{S^+}{F_s}{S^+}{K'}{1} \SixJ{N}{K}{I}{K'}{N{'}}{1} \SixJ{N^+}{N}{\ell}{N'}{N^{+}{'}}{1} 
\end{equation*}
\begin{equation}
\times \sqrt{(2K+1)(2K'+1)(2N+1)(2N'+1)(2N^++1)N^+(N^++1)(2S^++1)S^+(S^++1)}.
\label{eq:Hhfs8}
\end{equation}
\end{widetext}
An equivalent expression is obtained for the hyperfine coupling term $b_F (\bm{I} \cdot \bm{S^+})$:
\begin{widetext}
\begin{equation*}
b_F \braket{(\ell N^+)N(I)K(S^+)F_s(s)FM_F| \bm{T^1}(I) \cdot \bm{T^1}(S^+)|(\ell ' N^{+}{'})N'(I')K'(S^{+}){'}F_s'(s')F'M_F'} 
\end{equation*}
\begin{equation*}
=b_F \delta_{M_F M_F'}\delta_{FF'} \delta_{F_sF_s'} \delta_{NN'}  \delta_{II'} \delta_{S^+ S^{+}{'}}  (-1)^{K' + S^++F_s +N + I' + K +1 }  \SixJ{K}{S^+}{F_s}{S^{+}{'}}{K'}{1}  \SixJ{I}{K}{N}{K'}{I{'}}{1} 
\end{equation*}
\begin{equation}
\times \sqrt{(2K+1)(2K'+1)(2I+1)I(I+1)(2S^++1)S^+(S^++1)}.
\label{eq:Hhfs9}
\end{equation}
\end{widetext}
The full derivations of Eqs. \eqref{eq:Hhfs8} and \eqref{eq:Hhfs9} are given in Eqs. (SI) and (SII) of the Supplemental Material \cite{supplemental_doran26a}.

In the case of nonpenetrating Rydberg states, the magnetic spin-orbit and spin-spin interactions can be expressed using the Breit-Bethe approximation \cite{macadam75a,hessels87a, snow03a} as 
\begin{equation*}
 \hat{H}_\textrm{spin}^{\textrm{BB}} = \frac{e^2}{4\pi\epsilon_0} \left[ \frac{\alpha^2}{2} \frac{1}{r^3}\bm{\ell} \cdot \bm{s} - \alpha^2  \frac{1}{r^3}\bm{\ell} \cdot \bm{S^+} \right.  
\end{equation*}
\begin{equation*}
 \left.- \alpha^2 \frac{\bm{S^+} \cdot (1-3 \hat{r} \hat{r}) \cdot \bm{s}}{r^3} \right]
\end{equation*}
\begin{equation}
=\hat{H}_{\textrm{so}} + \hat{H}_{\textrm{oso}} + \hat{H}_{\textrm{ss}},
\label{eq:Hso_blume_processed}
\end{equation}
where $\alpha$ is the fine-structure constant and $r$ is the distance between the Rydberg electron and the ion-core center of charge. For systems where the electron in the ion core has no orbital angular momentum, Eq. \eqref{eq:Hso_blume_processed} is obtained starting from the Breit-Pauli Hamiltonian \cite{bethe57a, blume62a} and treating the ion core as a point charge. The last term, $\hat{H}_{\textrm{ss}}$, is the spin-spin interaction and can be neglected for high-$\ell$ Rydberg states \cite{snow03a, lundeen05a}. The first two terms represent the interactions of the orbital angular momentum of the Rydberg electron with its own spin (spin-orbit $\hat{H}_{\textrm{so}}$), and with the spin of the ion core electron (other-spin$-$orbit  $\hat{H}_{\textrm{oso}}$). For conciseness only the final expressions are given below and full derivations are presented in Eqs. (SIII-IV) of the Supplemental Material \cite{supplemental_doran26a}.

\begin{widetext}
\begin{equation*}
\bra{nv^+} \braket{(\ell N^+)N(I)K(S^+)F_s(s)FM_F|\hat{H}_\textrm{spin}^{\textrm{BB}} |(\ell' N^{+}{'})N'(I')K'(S^{+}{'})F_s'(s')F'M_F'} \ket{n'v^{+}{'}}
\end{equation*}
\begin{equation*}
= \frac{e^2}{4\pi\epsilon_0} \frac{\alpha^2}{2} \left<\frac{1}{r^3}\right> \delta_{M_F M_F'}\delta_{FF'}\delta_{S^+S^{+}{'}} \delta_{II'} \delta_{N^+N^{+}{'}} \delta_{\ell \ell'} \delta_{ss'} (-1)^{F_s' + s + F+ K + S^+ + F_s + N + I + K'  + \ell + N^+ + N'+1}
\end{equation*}
\begin{equation*}
\times \SixJ{F_s}{s}{F}{s'}{F_s'}{1}  \SixJ{K}{F_s}{S^+}{F_s'}{K'}{1} \SixJ{N}{K}{I}{K'}{N'}{1}  \SixJ{\ell}{N}{N^+}{N'}{\ell'}{1} 
\end{equation*}
\begin{equation*}
\times \sqrt{(2F_s+1)(2F_s'+1)(2K+1)(2K'+1)(2N+1)(2N'+1)(2\ell+1)\ell(\ell+1)(2s+1)s(s+1)}
\end{equation*}
\begin{equation*}
- \frac{e^2}{4\pi\epsilon_0} \alpha^2 \left<\frac{1}{r^3}\right>\delta_{M_F M_F'}\delta_{FF'} \delta_{F_sF_s'} \delta_{II'} \delta_{N^+ N^{+}{'}}  \delta_{\ell \ell'} \delta_{S^+S^{+}{'}}(-1)^{K' + S^++F_s +  N + I + K' + \ell + N^+ + N' } 
\end{equation*}
\begin{equation*}
\times \SixJ{K}{S^+}{F_s}{S^{+}{'}}{K'}{1} \SixJ{N}{K}{I}{K'}{N'}{1} \SixJ{\ell}{N}{N^+}{N'}{\ell'}{1} 
\end{equation*}
\begin{equation}
\times \sqrt{(2K+1)(2K'+1)(2N+1)(2N'+1)(2S^++1)S^+(S^++1)(2\ell+1)\ell(\ell+1)}.
\label{eq:Hso}
\end{equation}
\end{widetext}
The values of $<1/r^3>$ are obtained by averaging over the radial wavefunctions of appropriate quantum defect obtained numerically using the Numerov algorithm \cite{hoelsch22a}.

The magnitude of the exchange interaction in low-$n$, high-$\ell$ Rydberg states of molecular hydrogen has previously been investigated experimentally. Sturrus \textit{et al.} have reported a value of 500 kHz for the exchange interaction of the 10g state \cite{sturrus86a}, and Jungen \textit{et al.} \cite{jungen90a} estimated an upper bound of 0.005 cm$^{-1}$ (150 MHz) for the 5g state. Based on an $n^{-3}$ scaling, a conservative upper limit of 800 kHz is obtained for 30g. Using similar estimations, the exchange interaction for $\ell \geq 5$ can be safely neglected for principal quantum numbers $n \geq 30$.

\begin{figure*}
	{\includegraphics[trim=2cm 17cm 2cm 5cm, clip=true, width=0.95\linewidth]{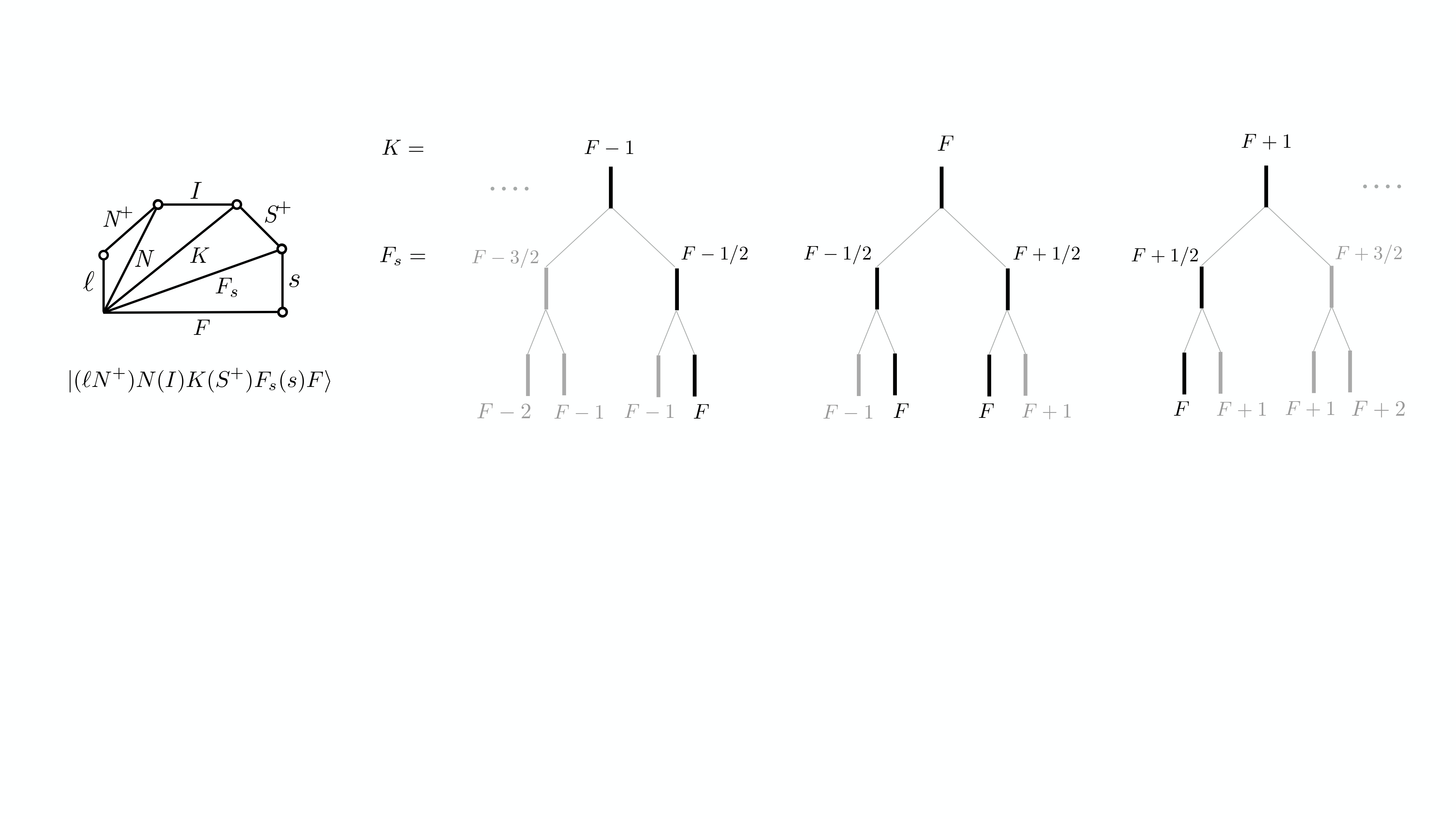}}
      \caption{Angular-momentum-coupling diagram in the $\ket{(\ell N^+)N(I)K(S^+)F_s(s)F}$ basis set (left) and subset of four interacting states with the total angular momentum $F$ (right).}  
              \label{fig:angmom_polmol}	
\end{figure*}

The energies of high-$\ell$ Rydberg states are obtained by diagonalizing the matrix $\hat{H}_{0, \ell \geq4}$ [Eq. \eqref{eq:H_LR_tot}] in the $\ket{n v^+} \ket{(\ell N^+)N(I)K(S^+)F_s(s)FM_F}$ basis. In the absence of an external field, $F$ is a good quantum number and all $M_F$ states are degenerate. Because the interactions considered above [see Eqs. \eqref{eq:HCoulomb4}, \eqref{eq:Hpol4}, \eqref{eq:Hhfs8}, \eqref{eq:Hhfs9}, \eqref{eq:Hso}] imply the selection rules $\Delta \ell = 0$, $\Delta N^+ = 0$ and $\Delta I = 0$, $\hat{H}_0$ is block-diagonal and each submatrix of states with the same $\ell, N^+, I, F$ can be diagonalized separately. The maximal dimension of such a submatrix is 4$\times$4, corresponding to $ (F-1) \leq |\vec{\ell} + \vec{N}^+ + \vec{I}| \leq (F+1) $, as illustrated by the angular-momentum-coupling diagram shown in Fig. \ref{fig:angmom_polmol}. 

\begin{figure*}
    \centering
     {\includegraphics[trim=6cm 2.5cm 6cm 1.2cm, clip=true, width=\linewidth]{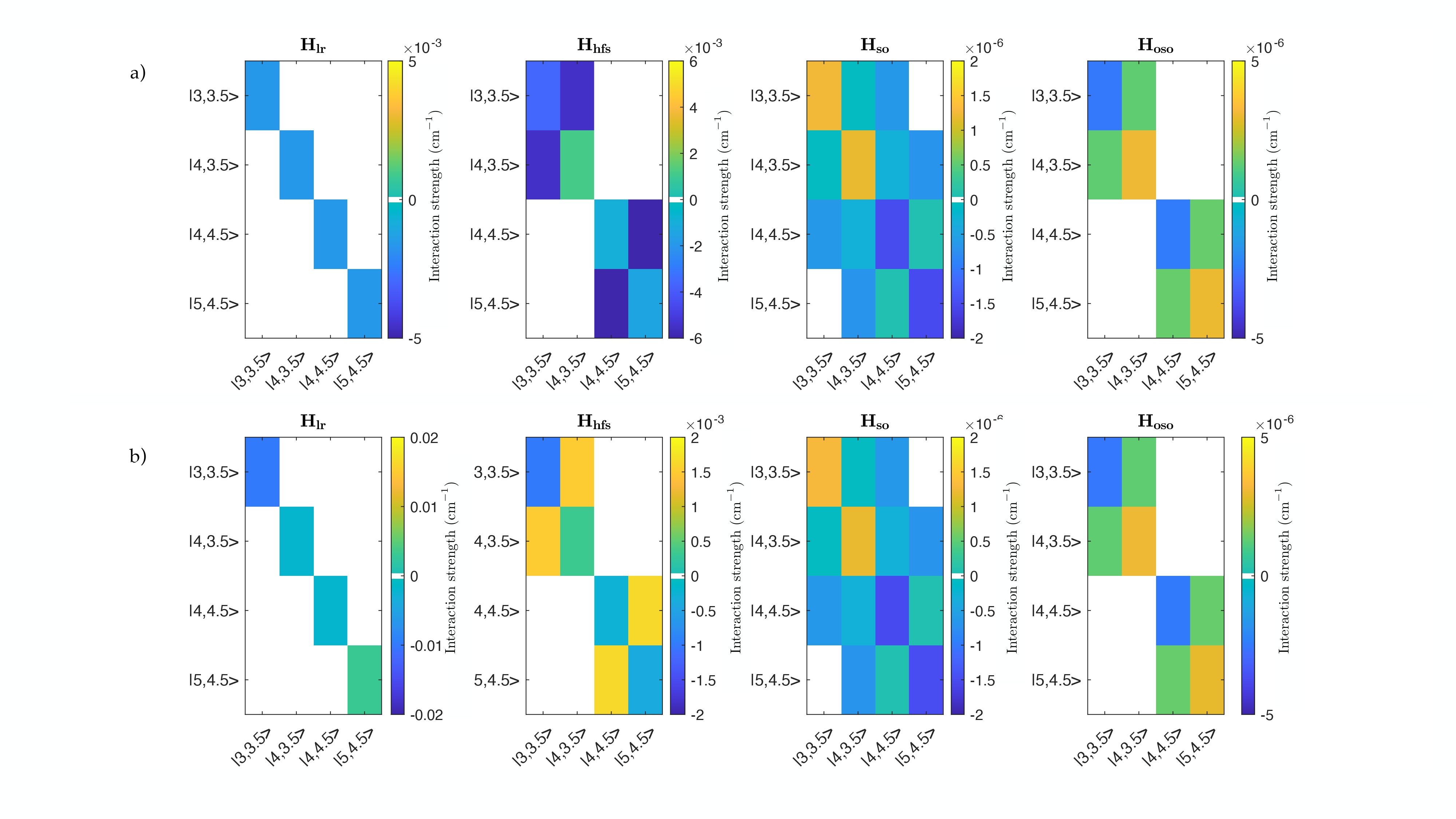}}
     \caption{4 $\times$4 matrices showing the electrostatic long-range ($\hat{H}_\textrm{lr}$), hyperfine ($\hat{H}_\textrm{hfs}$), spin-orbit ($\hat{H}_\textrm{so}$) and other-spin$-$orbit ($\hat{H}_\textrm{oso}$) interactions between $\ket{n v^+} \ket{(\ell N^+)N(I)K(S^+)F_s(s)F}$ states with $n=40, v^+=1, \ell=5, F=3$, for a) ortho-D$_2$ ($I=2, N^+=0$) and b) para-H$_2$ ($I =0, N^+=2$). The center of the color scale (corresponding to zero coupling) is set to white. The basis states are labeled as $\ket{K,F_s}$, corresponding to Fig. \ref{fig:angmom_polmol}.}
    \label{fig:relstr}
\end{figure*}

\begin{figure*}
    \centering
     {\includegraphics[trim=2.5cm 0cm 2cm 0cm, clip=true, width=0.97\linewidth]{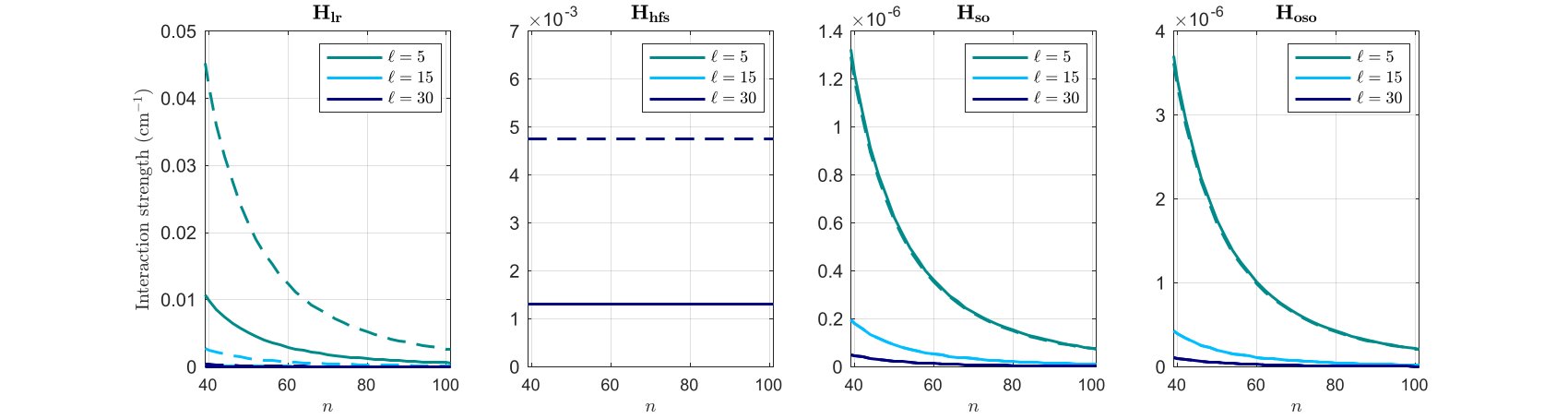}}
     \caption{Evolution of electrostatic long-range, hyperfine, spin-orbit and other-spin$-$orbit interactions in the range $n=40-100$, for $\ell = 5, 15, 30$ and $v^+=1$, plotted as the diagonal matrix element with $K=\ell - 2$, $F_s=K-1/2 = \ell-5/2$, $F=K=\ell-2$. The results for ortho-D$_2$ ($I=2, N^+=0$) are drawn as dashed lines and those for para-H$_2$ ($I=0, N^+=2$) as solid lines.}
    \label{fig:relstr_magnitudes}
\end{figure*}

\begin{figure*}
    \centering
		{\includegraphics[trim=0cm 13.5cm 0cm 1.5cm, clip=true, width=\linewidth]{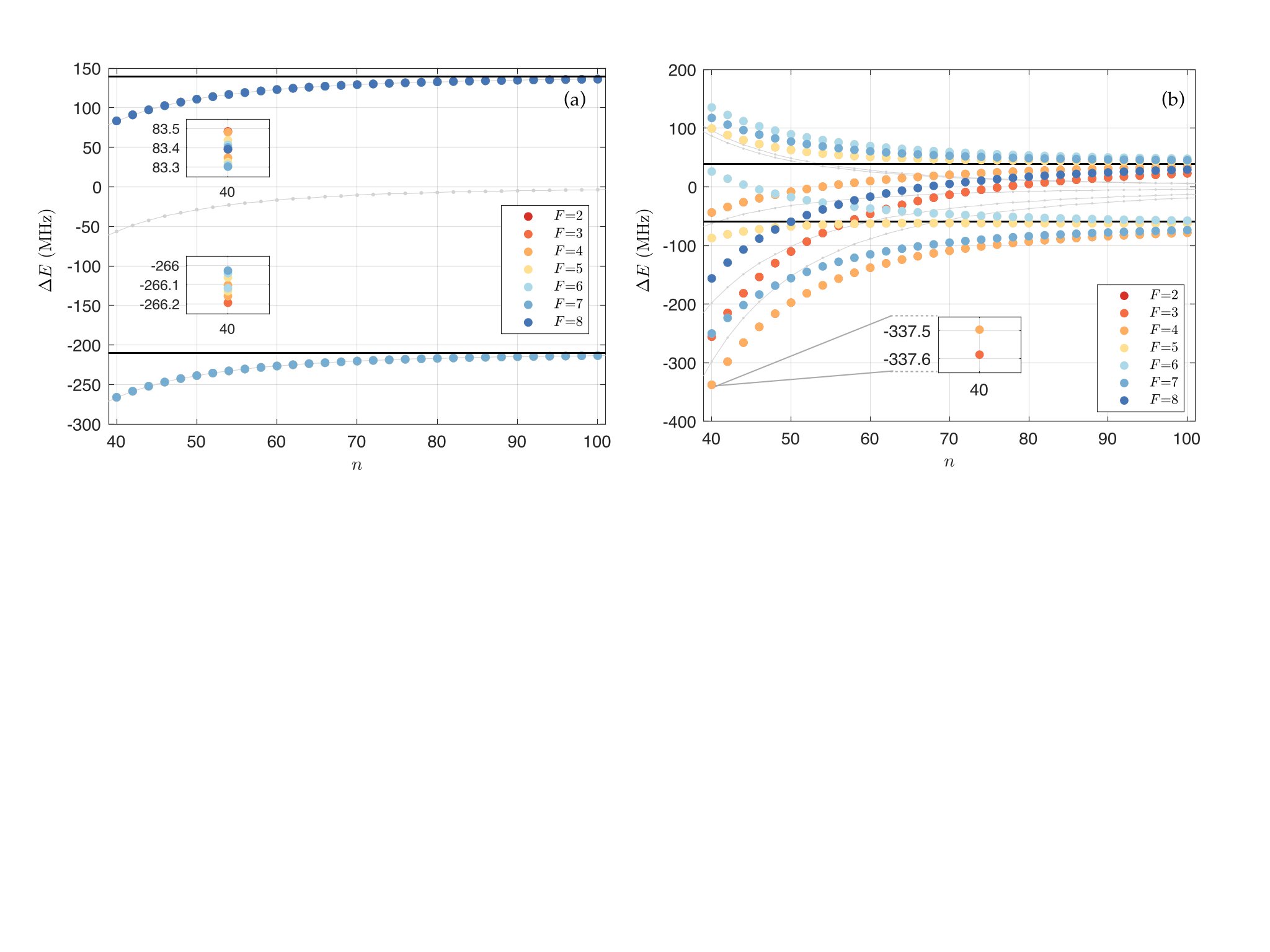}}
      \caption{Evolution of energy-level structure of $\ell=5$ Rydberg states of (a) ortho-D$_2$, $I = 2, v^+=1, N^+=0$ and (b) para-H$_2$, $I = 0, v^+=1, N^+=2$, as a function of $n$. The energy levels are referenced to $-R_{\textrm{D}_2 (\textrm{H}_2)} / n^2$ and the black horizontal lines indicate the ionic hyperfine (spin-rotation) levels. The light grey lines show the predicted level energies neglecting all spins (i.e., corresponding to the spin-independent long-range interaction model). The insets centered around $n=40$ show further splittings, in the range of 0.1-0.2 MHz, caused by spin-orbit couplings.}
    \label{fig:nevol}
\end{figure*}

Figure \ref{fig:relstr} shows the overall structure of these submatrices for $n=40$, $v^+=1$, $\ell=5$, in the case of (a) ortho-D$_2$ with $I=2, N^+=0$ and (b) para-H$_2$ with $I=0, N^+=2$. The elements of $\hat{H}_\textrm{Coulomb}$ only contribute constant diagonal elements in the chosen basis set [see Eq. \eqref{eq:HCoulomb4}] and are not represented in Fig. \ref{fig:relstr}. Whereas $\hat{H}_\textrm{lr}$ only has diagonal elements in the chosen basis set, $\hat{H}_\textrm{hfs}$ and $\hat{H}_\textrm{oso}$ are block-diagonal, because they conserve $F_s$, and $\hat{H}_\textrm{so}$ mixes all states. The matrix elements of $\hat{H}_\textrm{lr}$ (spin-independent) and $\hat{H}_\textrm{hfs}$ are comparable in magnitude, and those of the spin-orbit interactions $\hat{H}_\textrm{so}$ and $\hat{H}_\textrm{oso}$ are both approximately three orders of magnitude weaker. The energy-level structure of high-$\ell$ Rydberg states is therefore expected to depend on the interplay between the spin-independent long-range interaction (quadrupole and dipole polarizability) and the interaction term $\bm{I \cdot S^+}$ in the case of ortho-D$_2$ [Eq. \eqref{eq:Hhfs9}] and $\bm{N^+ \cdot S^+}$ in the case of para-H$_2$ [Eq. \eqref{eq:Hhfs8}]. In the case of para-H$_2$ with $N^+=2$, the electrostatic interaction between $\vec{\ell}$ and $\vec{N}^+$ splits the energy levels of different $K$ value ($\vec{K} = \vec{N} = \vec{\ell} + \vec{N}^+$). Consequently, the diagonal elements of $\hat{H}_\textrm{lr}$ are not equal. In the case of ortho-D$_2$ with $N^+=0$, no such splittings occur and the elements of $\hat{H}_\textrm{lr}$ are diagonal and identical.

To illustrate the evolution of the different interactions discussed above with $n$ and $\ell$, Fig. \ref{fig:relstr_magnitudes} shows the magnitude of a typical diagonal matrix element, chosen here to have $K=\ell - 2$, $F_s=K-1/2 = \ell-5/2$, $F=K=\ell-2$, as a function of $n$, for three different values of $\ell$. Similar behaviors are observed for each of the four types of interactions in ortho-D$_2$ (dashed lines) and para-H$_2$ (solid lines). The second panel depicts the hyperfine interaction, which is related to the angular momenta in the ion core only and therefore does not depend on $n$ or $\ell$. The strength of both long-range electrostatic and spin-orbit interactions decreases with $n$ (following an $n^{-3}$ scaling) and $\ell$, as expected. The elements of $\hat{H}_\textrm{so}$ and $\hat{H}_\textrm{oso}$ are more than three orders of magnitude weaker than those of $\hat{H}_\textrm{lr}$ and $\hat{H}_\textrm{hfs}$, as mentioned above in the discussion of Fig. \ref{fig:relstr}.

The energy-level structures are plotted for $\ell=5$ in Fig. \ref{fig:nevol}a for ortho-D$_2$ ($I=2, N^+=0$) and in Fig. \ref{fig:nevol}b for para-H$_2$ ($I=0, N^+=2$), using different colors for states with different values of the total angular momentum $F$. The light grey lines are displayed at the level energies predicted when neglecting spins, corresponding to $\hat{H}_\textrm{lr}$ in Eq. \eqref{eq:Hpol4}. In the case of ortho-D$_2$, the energy levels form two distinct groups separated by approximately 350 MHz, corresponding to the two possible values of $F^+ (= G^+)$ of 1.5 and 2.5. This situation corresponds to the angular-momentum-coupling schemes $\ket{1}$ and $\ket{3}$ in Fig. \ref{fig:basis_sets}. To an excellent approximation, these two groups of energy levels with $F^+=1.5$ and 2.5 are displaced from the spin-independent energy levels (light grey line) by the hyperfine energies of the D$_2^+$ ($v^+=1, N^+=0$) ion (shown as horizontal black lines). The two insets centered around $n=40$ show that the energy splittings caused by spin-orbit interactions are much smaller, on the order of 200 kHz for both groups of $F^+$ levels.

The energy-level pattern in the case of para-H$_2$ is very different and changes rapidly with $n$: at $n$ values around 40, the structure resembles that predicted by a spin-independent calculation (light grey lines), where $N$ ($ \vec{N} = \vec{\ell} + \vec{N}^+$) is a good quantum number, corresponding to the second basis set in Fig. \ref{fig:basis_sets}. With increasing $n$, $N$ ceases to be a good quantum number. In the high-$n$ ($\approx$ 100) limit, the states form two groups of near-degenerate levels separated by the spin-rotation splitting of the $N^+=2$ ionic levels (horizontal black lines). The energy levels in these two groups can be labeled by distinct $F^+ (= J^+)$ values of 1.5 and 2.5, corresponding to the coupling schemes $\ket{1}$ and $\ket{3}$ in Fig. \ref{fig:basis_sets}. As in the case of D$_2$, the spin-orbit interactions cause much smaller splittings, as indicated in the inset for $n=40$. $\hat{H}_\textrm{lr}$ and $\hat{H}_\textrm{hfs}$ are much larger than $\hat{H}_\textrm{so}$ and $\hat{H}_\textrm{oso}$ (see Fig. \ref{fig:relstr}) and the energy-level structure is dominated by the electrostatic long-range and hyperfine interactions. 

The fundamentally different behaviors of the level structures of ortho-D$_2$ and para-H$_2$ stem from the absence of electrostatic coupling between $\vec{\ell}$ and $\vec{N}^+$ for ortho-D$_2$ ($N^+=0$). As a result, the $n$-independent hyperfine interaction in the D$_2^+$ ion core fully determines the level structure of the Rydberg states. In para-H$_2$ with $N^+=2$ on the other hand, the level structure is more complex and reflects the interplay between the $n^{-3}$-dependent electrostatic coupling ($\vec{N} = \vec{\ell} + \vec{N}^+$) and the spin-rotation coupling of the ion core. In both cases, including the effects of the fine and hyperfine interactions is essential to correctly describe the convergence of the Rydberg series in the high-$n$ limit. The accuracy of the calculations presented in this section was validated by comparison with MQDT calculations performed for g and h states using quantum defects obtained from a long-range model including higher-order corrections \cite{jungen89a, jungen90a, jungen25p} (see Section \ref{sec:errors}). 

\section{Calculation of the Stark effect}
\label{sec:stark}

The matrix elements of the operator $e \mathcal{F} \hat{z}$ describing the effect of the electric field are calculated using the $ \ket {(\ell s)j \left[ (S^+ N^+)J^+(I) \right]F^+F M_F}$ basis ($\ket 3$ in Fig. \ref{fig:basis_sets}) after sequential decoupling of the angular momenta. The matrix elements of the $\hat{z}$ operator in this basis are given by

\begin{widetext}
\begin{equation*}
\bra{nv^+} \braket{(\ell s)j (S^+ N^+)J^+(I)F^+F{M_F}|\hat{z}|(\ell' s')j' (S^{+}{'} N^{+}{'})J^{+}{'} (I')F^{+}{'}F'{M_F'}} \ket{n'v^{+}{'}}
\end{equation*}
\begin{equation*}
=(-1)^{F-M_F}\ThreeJ{F}{1}{F'}{-M_F}{0}{M_{F}'}  \bra{nv^+} \braket{ (\ell s)j (S^+ N^+)J^+(I)F^+F||r||(\ell' s')j' (S^{+}{'} N^{+}{'})J^{+}{'} (I')F^{+}{'}F'} \ket{n' v^{+}{'}}
\end{equation*}
\begin{equation*}
=\delta_{F^+ F^{+}{'}} (-1)^{F - M_F + j+F^+ + F' + 1} \sqrt{(2F+1)(2F'+1)}\ThreeJ{F}{1}{F'}{-M_F}{0}{M_{F}'}  \SixJ{j}{F}{F^+}{F'}{j'}{1} \bra{nv^+}\braket{(\ell s)j||r||(\ell' s')j'}  \ket{n' v^{+}{'}}
\end{equation*}
\begin{equation*}
=\delta_{F^+ F^{+}{'}} \delta_{ss'} (-1)^{F - M_F + j+F^+ + F' + 1+ \ell + s + j' + 1} \sqrt{(2F+1)(2F'+1)(2j+1)(2j'+1)} 
\end{equation*}
\begin{equation*}
\times \ThreeJ{F}{1}{F'}{-M_F}{0}{M_{F}'}  \SixJ{j}{F}{F^+}{F'}{j'}{1}  \SixJ{\ell}{j}{s}{j'}{\ell'}{1}  \braket{n v^+\ell ||r||n' v^{+}{'}\ell'}   
\end{equation*}
\begin{equation*}
=\delta_{F^+ F^{+}{'}} \delta_{ss'} (-1)^{F - M_F + j+F^+ + F' + \ell + s + j'} \sqrt{(2F+1)(2F'+1)(2j+1)(2j'+1)}\ThreeJ{F}{1}{F'}{-M_F}{0}{M_{F}'}  \SixJ{j}{F}{F^+}{F'}{j'}{1}  \SixJ{\ell}{j}{s}{j'}{\ell'}{1}
\end{equation*}
\begin{equation}
\times (-1)^{\ell}\ThreeJ{\ell}{1}{\ell'}{0}{0}{0}\sqrt{(2 \ell+1)(2 \ell'+1)}\braket{nv^+ \ell |r|n' v^{+}{'}\ell'}\,.
\label{eq:rstark6}
\end{equation}
\end{widetext}
The radial integrals $\braket{nv^+ \ell |r|n' v^{+}{'}\ell'}$ are calculated numerically using the Numerov algorithm \cite{hoelsch22a}. The angular momenta of the ion core ($\vec{F}^+ = \vec{S}^+ + \vec{N}^+ + \vec{I}$) have to be coupled separately from those of the Rydberg electron ($\vec{j} = \vec{\ell} + \vec{s}$), because the binding energies of the basis states need to be specified relative to a given rovibrational/hyperfine ionic level for the Numerov algorithm.

\begin{figure}
    \centering
     {\includegraphics[trim=0cm 0cm 0cm 0cm, clip=true, width=\linewidth]{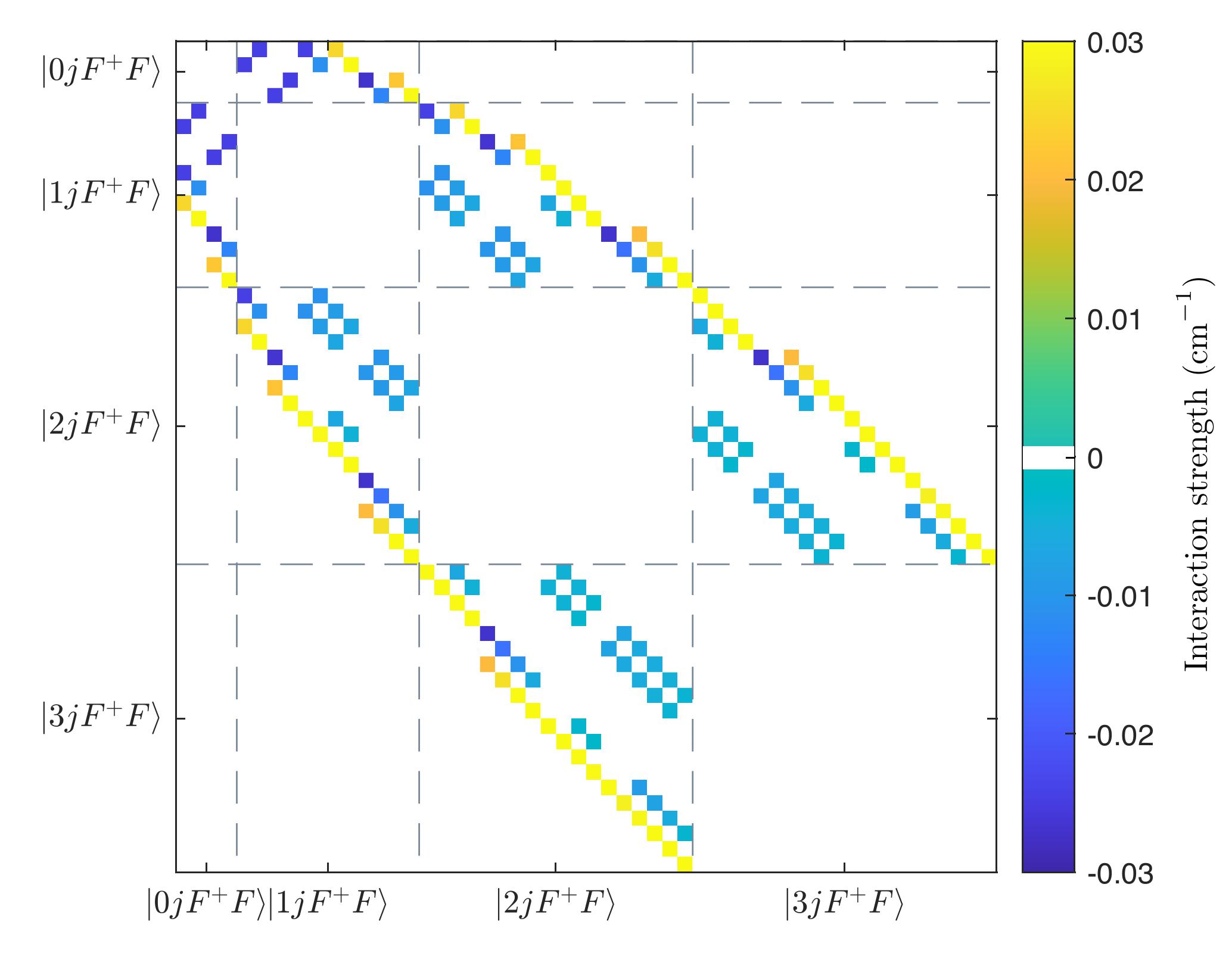}}
      \caption{Stark matrix (subset for $\ell=0-3$) for para-H$_2$, $I=0, N^+=2$, $n = 34, v^+=1, M_F=0$, and an electric field strength of 1 V/cm. The basis functions are labeled by $\ket{\ell jF^+F}$. }
    \label{fig:matrix_stark_ph2}
\end{figure}

Fig. \ref{fig:matrix_stark_ph2} illustrates the first few entries ($\ell = 0-3$) in the matrix representation of $e \mathcal{F} \hat{z}$ in the $\ket {(\ell s)j \left[ (S^+ N^+)J^+(I) \right] F^+FM_F}$ basis for para-H$_2$ ($I=0, N^+=2$), $n=34, v^+=1, M_F=0$ and an electric field strength of  1 V/cm. As in Fig. \ref{fig:matrices_Hzero_oD2_pH2}, only the value of $\ell$ is explicitly given in the labels (the complete labels with the values of $j, F^+$ and $F$ are given in Fig. S3 in the Supplemental Material \cite{supplemental_doran26a}). The matrix of $e \mathcal{F} \hat{z}$ is symmetric and block-diagonal, with submatrices of nonzero elements corresponding to the selection rule $\Delta \ell = \pm1$ indicated by the dashed rectangular frames. At an electric field strength of 1 V/cm, the magnitude of the off-diagonal, field-induced interactions is on the order of 0.03 cm$^{-1}$ (1 GHz). The matrix of $e \mathcal{F} \hat{z}$ in the case of ortho-D$_2$, $I=2, N^+=0$ (given in Fig.S4 in the Supplemental Material \cite{supplemental_doran26a}) is almost identical, because there is only a dependence on $F^+$ ($\vec{F}^+ = \vec{S}^+ + \vec{N}^+ + \vec{I}$) in Eq. \eqref{eq:rstark6}, and not on $N^+$ or $I$ individually. The matrix elements for para-H$_2$ and ortho-D$_2$ only differ slightly in their numerical values, because of the different values of the radial integrals $\braket{nv^+ \ell |r|n' v^{+}{'}\ell'}$.

\section{Intensity model}
\label{sec:intensity_model}

The eigenvectors of the total Hamiltonian matrix [Eq. \eqref{eq:htotstark_intro}] at a given value of the electric field strength are linear combinations of $ \ket {(\ell s)j (S^+ N^+)J^+(I)F^+F M_F}$ basis vectors. However, the calculation of transition intensities requires a Hund's case b) treatment $ \ket {(\ell N^+)N_\Lambda (sS^+)S J (I) F M_F}$ [basis set $\ket{4}$ in Figure \ref{fig:basis_sets}, where the total electron spin $S$ ($\vec{S} = \vec{s} + \vec{S}^+$) and $\Lambda$ are defined] because absorption takes place at short range, where $S$ and $\Lambda$ are good quantum numbers. When the ion core is in a $\Sigma$ electronic state, $\Lambda = \lambda$, which is also the quantum number for the projection of the total angular momentum without spins $N$ ($\vec{N} = \vec{\ell} + \vec{N}^+$) onto the internuclear axis. The frame transformation between these two basis sets is given in Section \ref{sec:frametrafos}. The eigenvectors $\ket{\phi}$ of the Rydberg-Stark states can then be expressed as 
\begin{equation}
\ket{\phi} = \sum_{m} c_{m} \ket { \left[(N_{\Lambda}S) J (I) F \right]_m {M_F} },
\label{eq:hundb}
\end{equation}
where $m$ denotes the different basis states in the $ \ket {(\ell N^+)N_\Lambda (sS^+)S J (I) F M_F}$ basis. $M_F$ is a good quantum number for the Rydberg-Stark states in a homogeneous electric field, i.e., $(M_F)_m = M_F$.

The electric-dipole interaction is expressed as the negative vector product of the electric field vector $\vec{E}$ of the electromagnetic radiation and the electric dipole operator $\vec{\mu_\textrm{e}}$ \cite{brown03a} :
\begin{equation*}
\hat{H}^{\textrm{dip}} = - \vec{E_{}} \cdot \vec{\mu_{\textrm{e}}} = - \bm{T^{(1)}} (\vec{E}) \cdot \bm{T^{(1)}} (\vec{\mu_{\textrm{e}}}) 
\end{equation*}
\begin{equation}
= - \sum_{p=-1}^{1} (-1)^p T_p^{(1)} (\vec{E}) T_{-p}^{(1)} (\vec{\mu_{\textrm{e}}}),
\label{eq:def_Hdip}
\end{equation}
where the index $p$ labels the components of the spherical tensors in the laboratory-fixed frame. Expressing the dipole operator in the molecule-fixed coordinate system using Wigner rotation matrices \cite{zare88a}
\begin{equation*}
T_{p}^{(1)} (\vec{\mu_{\textrm{e}}}) = \sum_{q} D_{pq}^{(1)}  (\omega)^* \ T_q ^{(1)} (\vec{\mu_{\textrm{e}}}),
\end{equation*}
where the label $q$ denotes the components of $\vec{\mu_{\textrm{e}}}$ in the molecule-fixed frame, yields
\begin{equation}
\hat{H}^{\textrm{dip}} = - \sum_{pq} (-1)^p  \ T_p^{(1)} (\vec{E}) \ D_{-pq}^{(1)}  (\omega)^* \ T_q ^{(1)} (\vec{\mu_{\textrm{e}}}).
\label{eq:worked_Hdip}
\end{equation}
Evaluating the elements of $D_{-pq}^{(1)}  (\omega)^*$ in the $ \ket {(\ell N^+)N_\Lambda (sS^+)S J (I) F M_F}$ basis set (designated as $ \ket {N_\Lambda (sS^+)S J (I) F M_F}$ below for conciseness) results in (see full derivation in SV of the Supplemental Material \cite{supplemental_doran26a}) 
\begin{widetext}
\begin{equation*}
\braket{ (N_{\Lambda}S) J (I) F {M_F}|D_{-pq}^{(1)}  (\omega)^* | (N'_{\Lambda'}S') J' (I') F' {M_F}'}  
\end{equation*}
\begin{equation*}
= \delta_{I I'} \delta_{SS'} (-1)^{F - M_F + J + I + F' + N + S + J'  + N - \Lambda} \ThreeJ{F}{1}{F'}{-M_F}{-p}{M_{F}'}  \SixJ{J}{F}{I}{F'}{J'}{1}  \SixJ{N}{J}{S}{J'}{N'}{1} \ThreeJ{N}{1}{N'}{-\Lambda}{q}{\Lambda}  
\label{eq:matel_seven}
\end{equation*}
\begin{equation}
\times \sqrt{(2F+1)(2F'+1)(2J+1)(2J'+1)(2N+1)(2N'+1)},
\label{eq:matel_transi}
\end{equation}
\end{widetext}
where the values of $p$ determine the polarization of the radiation field with respect to the direction of the static field ($z$-axis of the laboratory frame): $p = 0$ for parallel, $p= \pm 1$ for perpendicular and $p = +1 (-1) $ for circular polarizations. Similarly, $q= 0$ and $q= \pm 1$ designate parallel and perpendicular transitions in the molecule-fixed frame. 

We consider an experimental scheme in which the Rydberg-Stark states $\ket{\phi_4}$ are accessed from the ground state $\ket{\phi_1}$ in a resonant three-photon-excitation sequence through the intermediate states $\ket{\phi_2}$ and $\ket{\phi_3}$, which are to an excellent approximation described by Hund's case (b) basis states $\ket { (\ell N^+)N_{\Lambda}(S) J (I) F {M_F} }$. In the experiments suggested to test these calculations \cite{iodo_tbp} (see also discussion of Fig. \ref{fig:starkmaps} below), the transitions $\ket{\phi_1} \rightarrow \ket{\phi_2}$ and $\ket{\phi_2} \rightarrow \ket{\phi_3}$ are $\Sigma \rightarrow \Sigma$, such that $q=0$ in both cases. The Rydberg-Stark states $\ket{\phi_4}$ are superpositions of different $\lambda \leq \ell$ states [Eq. \eqref{eq:hundb}]. Consequently, starting from the $\ket{\phi_3}$ electronic state with $\Sigma$ character, the allowed transitions are $\Sigma \rightarrow \Sigma$ ($\Sigma \rightarrow \Pi$) for $q=0$ ($q= \pm 1$). The line strength for a three-photon (3ph) transition is
\begin{equation}
S = \sum_{M_{F_1}}\sum_{M_{F_4}}\left|\braket{ \phi_{1} |\hat{H}_{\textrm{3ph}}^{\textrm{dip}}|  \phi_{4}} \right| ^ 2
\label{eq:linestrength_gen}
\end{equation}
when starting from an initial state $\ket{\phi_1}$ with a given value of the total angular momentum $F_1$. Incoherent sums are carried out over the possible values of $M_{F_1}$ and $M_{F_4}$, where the value of $M_{F_4}$ is defined relative to the value of $M_{F_3}$ (corresponding to $\ket{\phi_3}$) by the polarization of the radiation field in the laboratory frame in the $\ket{\phi_3} \rightarrow \ket{\phi_4}$ step: $M_{F_4}$ = $M_{F_3}$ if $p_{34} = 0$ and $M_{F_4}$ = $M_{F_3} \pm 1 (+1,-1)$ if $p_{34} = \pm1 (+1,-1)$. A sum over $M_{F_4}$ is only needed for $p_{34} = \pm1$.

Using Eqs. \eqref{eq:worked_Hdip} and \eqref{eq:matel_transi}, we obtain
\begin{widetext}
\begin{equation*}
\braket{ \phi_{1} |\hat{H}_{\textrm{3ph}}^{\textrm{dip}}|  \phi_{4}} = \sum_{F_2,M_{F_2}} \sum_{F_3,M_{F_3}}  T_{p_{12}}^{(1)} (\vec{E}) \ \braket{  \left[ (\ell N^+) N_{\Lambda}(S) J (I) F M_F \right]_1 |T_{q_{12}} ^{(1)} (\vec{\mu_{\textrm{e}}})|   \left[ (\ell N^+) N_{\Lambda}(S) J (I) F M_F \right]_2} 
\end{equation*}
\begin{equation*}
\times \ T_{p_{23}}^{(1)} (\vec{E}) \ \braket{  \left[ (\ell N^+) N_{\Lambda}(S) J (I) F M_F \right]_2 |T_{q_{23}} ^{(1)} (\vec{\mu_{\textrm{e}}})|   \left[ (\ell N^+) N_{\Lambda}(S)J (I) F M_F \right]_3} 
\end{equation*}
\begin{equation*}
 \times  \sum_{m} \ T_{p_{3m}}^{(1)} (\vec{E}) \ \braket{  \left[ (\ell N^+) N_{\Lambda}(S) J (I) F M_F \right]_3 |T_{q_{3m}} ^{(1)} (\vec{\mu_{\textrm{e}}})| c_m \{  \left[ (\ell N^+) N_{\Lambda}(S) J (I) F \right]_m M_{F_4} \}} 
\end{equation*}
\begin{equation*}
\times \delta_{I_1I_2} \delta_{S_1S_2}  \delta_{I_2I_3} \delta_{S_2S_3} \delta_{I_3I_m} \delta_{S_3S_m} 
\end{equation*}
\begin{equation*}
\times (-1)^{p_{12} + F_1 - M_{F_1} + J_1 + I_1 + F_2  + S_1 + J_2 - \Lambda_1 + 1}  \sqrt{(2F_1+1)(2F_2+1)(2J_1+1)(2J_2+1)(2N_1+1)(2N_2+1)} 
\end{equation*}
\begin{equation*}
\times (-1)^{p_{23} + F_2 - M_{F_2} + J_2 + I_2 + F_3  + S_2 + J_3 - \Lambda_2 + 1}  \sqrt{(2F_2+1)(2F_3+1)(2J_2+1)(2J_3+1)(2N_2+1)(2N_3+1)} 
\end{equation*}
\begin{equation*}
\times (-1)^{p_{3m} + F_3 - M_{F_3} + J_3 + I_3 + F_m  + S_3 + J_m - \Lambda_3 + 1}  \sqrt{(2F_3+1)(2F_m+1)(2J_3+1)(2J_m+1)(2N_3+1)(2N_m+1)} \end{equation*}
\begin{equation*}
\times \ThreeJ{F_1}{1}{F_2}{-M_{F_1}}{-p_{12}}{M_{F_2}} \SixJ{J_1}{F_1}{I_1}{F_2}{J_2}{1}  \SixJ{N_1}{J_1}{S_1}{J_2}{N_2}{1} \ThreeJ{N_1}{1}{N_2}{-\Lambda_1}{q_{12}}{\Lambda_2}  
\end{equation*}
\begin{equation*}
\times \ThreeJ{F_2}{1}{F_3}{-M_{F_2}}{-p_{23}}{M_{F_3}} \SixJ{J_2}{F_2}{I_2}{F_3}{J_3}{1}  \SixJ{N_2}{J_2}{S_2}{J_3}{N_3}{1} \ThreeJ{N_2}{1}{N_3}{-\Lambda_2}{q_{23}}{\Lambda_3}  
\end{equation*}
\begin{equation}
\times \ThreeJ{F_3}{1}{F_m}{-M_{F_3}}{-p_{34}}{M_{F_4}} \SixJ{J_3}{F_3}{I_3}{F_m}{J_m}{1}  \SixJ{N_3}{J_3}{S_3}{J_m}{N_m}{1} \ThreeJ{N_3}{1}{N_m}{-\Lambda_3}{q_{3m}}{\Lambda_m}.
\label{eq:threeph}
\end{equation}
\end{widetext}

The coherent sums in Eq. \eqref{eq:threeph} span the possible combinations of $F_2$, $M_{F_2}$, $F_3$ and $M_{F_3}$ connecting the initial state with $F_1$ and $M_{F_1}$ to the final Rydberg-Stark states with $M_{F_4}$. As discussed above, $q_{12} = q_{23} = 0$, $q_{34} = 0, \pm1$, and $p_{12}, p_{23}$ and $p_{34}$ are defined by the polarizations of the radiation fields in the respective steps of the excitation scheme.

\begin{figure}
    \centering
		{\includegraphics[trim=0.5cm 0.5cm 24cm 0.5cm, clip=true, width=0.97\linewidth]{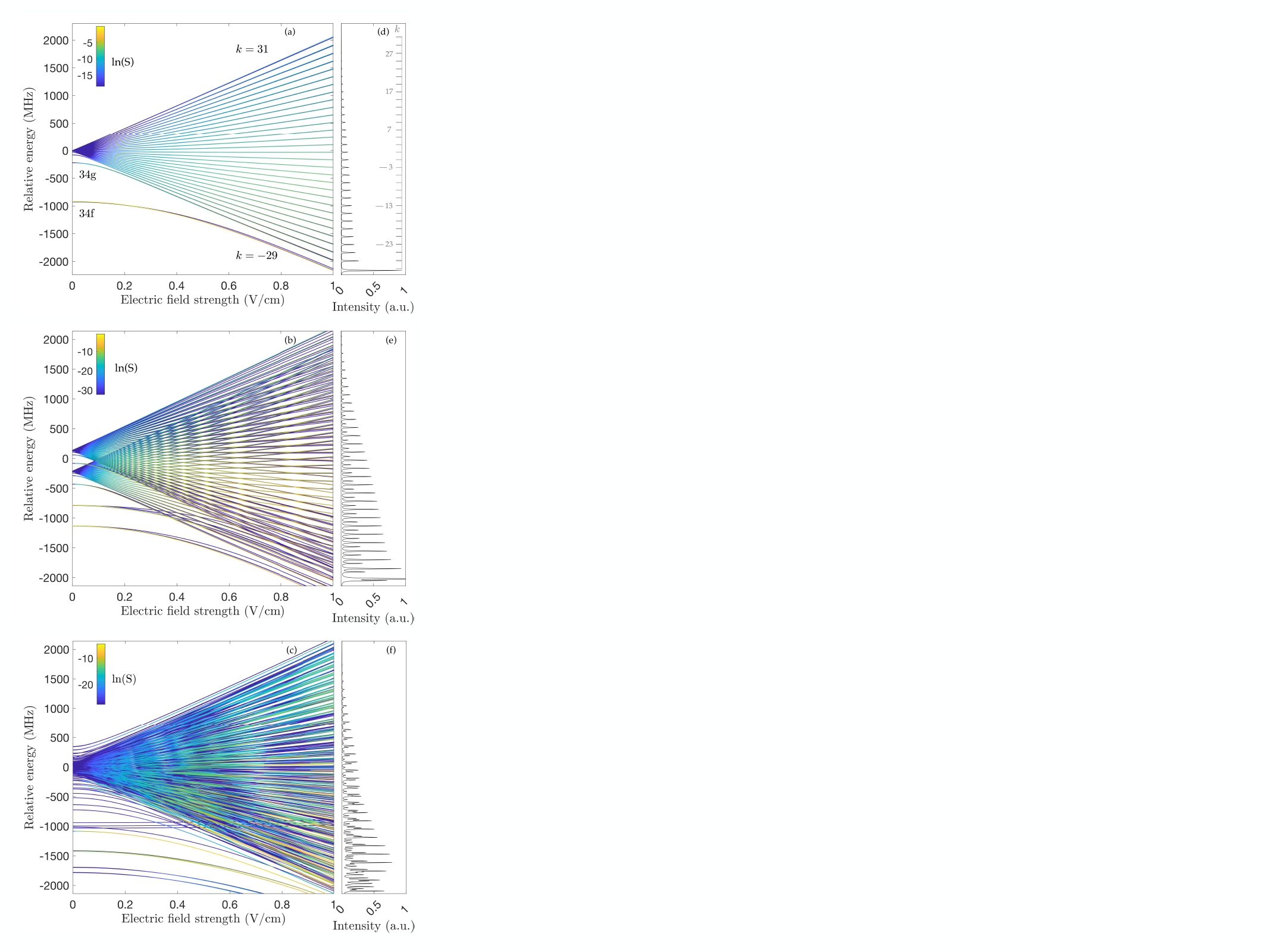}}
      \caption{Calculated Rydberg-Stark manifolds for $n=34, v^+=1$ \{(a) ortho-D$_2, I = 0, N^+=0$, (b) ortho-D$_2, I = 2, N^+=0$, both accessed with the multiphoton excitation sequence $\text{X}\,^1\Sigma_g^+(v=0, N=0) \rightarrow \text{B}\,^1\Sigma_u^+(v=4,N=1) \rightarrow \text{GK}\,^1\Sigma_g^+\, (v=2,N=2) \rightarrow n = 34 \,\,[\text{D}_2^+\,\text{X}{^{+}}\,^{2} \Sigma_g^+ (v^+=1, N^+=0) \,]$ and (c) para-H$_2, I = 0, N^+=2$, accessed with the multiphoton excitation sequence $\text{X}\,^1\Sigma_g^+(v=0, N=0) \rightarrow \text{B}\,^1\Sigma_u^+(v=4,N=1) \rightarrow \text{GK}\,^1\Sigma_g^+\, (v=2,N=0) \rightarrow n = 34 \,\,[\text{H}_2^+\,\text{X}{^{+}}\,^{2} \Sigma_g^+ (v^+=1, N^+=2) \,]$\} and corresponding spectra at 1 V/cm [panels (d-f)]. The logarithmic color scales indicate the predicted line strengths (S) corresponding to an experiment with all laser polarizations chosen parallel to the dc-electric field. The energy levels are given with respect to the spin-independent zero-quantum-defect positions $-R_{\textrm{D}_2(\textrm{H}_2)}/34^2$.}
    \label{fig:starkmaps}
\end{figure}

Fig. \ref{fig:starkmaps} shows calculated Rydberg-Stark manifolds for $n=34, v^+=1$, in (a) ortho-D$_2, I = 0, N^+=0$, (b) ortho-D$_2, I = 2, N^+=0$, and (c) para-H$_2, I = 0, N^+=2$ for varying electric field strengths between 0 and 1 V/cm. The corresponding spectra at 1 V/cm are displayed in panels (d)-(f) for $p_{12} = p_{23} = p_{34} = 0$, assuming a Lorentzian lineshape with a full width at half maximum of 5 MHz. The energy levels are referenced to the spin-independent zero-quantum-defect position $-R_{\textrm{D}_2(\textrm{H}_2)}/34^2$ with respect to the centroid of the hyperfine structure of the ionic level. The color scheme (logarithmic scale) indicates the line strengths calculated with Eqs. \eqref{eq:linestrength_gen} and \eqref{eq:threeph}, corresponding to the multiphoton excitation schemes $\text{X}\,^1\Sigma_g^+(v=0, N=0) \rightarrow \text{B}\,^1\Sigma_u^+(v=4,N=1) \rightarrow \text{GK}\,^1\Sigma_g^+\, (v=2,N=2) \rightarrow n = 34 \,\,[\text{D}_2^+\,\text{X}{^{+}}\,^{2} \Sigma_g^+ (v^+=1, N^+=0) \,]$ [for (a) and (b)] and $\text{X}\,^1\Sigma_g^+(v=0, N=0) \rightarrow \text{B}\,^1\Sigma_u^+(v=4,N=1) \rightarrow \text{GK}\,^1\Sigma_g^+\, (v=2,N=0) \rightarrow n = 34 \,\,[\text{H}_2^+\,\text{X}{^{+}}\,^{2} \Sigma_g^+ (v^+=1, N^+=2) \,]$ for (c), where in each step the polarization of the radiation field is parallel to the dc-field vector. Because the transitions from the GK state (of mostly 3d$\sigma$ character) to f Rydberg states are approximately two orders of magnitude stronger than to p Rydberg states \cite{hoelsch22a}, only contributions to Eq. \eqref{eq:threeph} from $ \ket{ \{ \left[ (\ell N^+) N_{\Lambda}(S) J (I) F \right]_m M_{F_4} \} }$ Rydberg-Stark states with $\ell_m$ = 3 are taken into account. Nonzero contributions to the line strengths are only obtained from states with $S_m$ = 0, because the ground $\ket{\phi_1}$ and intermediate electronic states $\ket{\phi_2}$ and $\ket{\phi_3}$ have $S=0$. We assume that the magnitudes of the dipole transition moments of $T_{q} ^{(1)} (\vec{\mu_{\textrm{e}}})$ are the same for $q=0,\pm1$.

The case of ortho-D$_2$ ($I=0, N^+=0$) in Fig. \ref{fig:starkmaps}a serves as a rotation- and spin-free reference and is equivalent to the situation encountered in para-H$_2$, $I=0, N^+=0$ \cite{hoelsch22a, doran24a}. Singlet and triplet states are not mixed in this case, and the singlet Rydberg-Stark states can be labeled by $k$. At zero-field, the high-$\ell$ states are centered around the zero-quantum-defect (ZQD) position, while the low-$\ell$ states with quantum defects significantly different from zero are displaced from the ZQD position by up to 1 GHz in the case of the 34f state. The s, p and d states corresponding to $k$ values of 33, $-33$ and $-31$, respectively, are displaced even further from the high-$\ell$ manifold and are located outside the displayed spectral range. At low values of the electric-field strengths ($<$ 0.5 V/cm), the 34f state is subject to a quadratic Stark shift and then merges with the linear Stark manifold at higher electric field strengths. 

In the presence of a nonzero nuclear spin [ortho-D$_2, I = 2, N^+=0$, panel (b)], two almost identical and independent linear Stark manifolds centered at approximately $-$210 MHz and 140 MHz [corresponding to the ionic hyperfine energy levels with $F^+=1.5$ and $F^+=2.5$ ($\vec{F}^+ = \vec{S}^+ + \vec{I}$), respectively] can be distinguished. These manifolds are almost identical to the Stark manifold of ortho-D$_2$ ($N^+=0, I=0$) in panel a). These observations imply that $F^+$ is a good quantum number in the presence of the electric field in this case, and that $k$ is still pattern-determining. The f and g states are displaced from the ZQD position at zero-field, and merge into the high-$\ell$ manifold at increasing values of the electric field strength. In Fig. \ref{fig:starkmaps}e) showing the calculated spectrum at 1 V/cm, the components of the two manifolds with $F^+=1.5$ and $F^+=2.5$ can be clearly distinguished. 

For para-H$_2$ [panels c) and f)], individual Stark manifolds can no longer be identified, i.e., $k$ is no longer pattern-determining. At zero-field, the energy levels span the entire range of $\pm$ 200 MHz around the spin-independent zero-quantum-defect position $-R_{\textrm{H}_2}/34^2$. In contrast to ortho-D$_2^+$ ($N^+=0, I=2$), states with different values of $F^+$ ($\vec{F}^+ = \vec{S}^+ + \vec{N}^+$ in this case) are now mixed (see also Fig. \ref{fig:nevol}b) by the electrostatic interaction ($\vec{N} = \vec{\ell} + \vec{N}^+$), which leads to a much more complex structure when $\ell$ is mixed by the electric field. As a result, neither the intensity distribution nor the line positions in the calculated spectrum (panel f) at 1 V/cm consistently exhibit distinct patterns reflecting the spin-rotation splitting between the $J^+=1.5$ and $J^+=2.5$ ($\vec{J}^+ = \vec{S}^+ + \vec{N}^+$) states of the H$_2^+$ ($N^+=2$) ion.

\section{Accuracy of calculations and sources of error}
\label{sec:errors}
Possible systematic errors in the calculated Stark shifts can be classified as related to either the field-free energies (Section \ref{sec:fieldfree}), or the field-induced energy shifts (Section \ref{sec:stark}). 

Errors in the field-free energies can be quantified in a sensitivity analysis, as explained in detail in Ref. \cite{hoelsch22a}. In the case of the low-$\ell$ Rydberg states, the uncertainties in the predicted field-free energies result from inaccuracies in the quantum defects used in the MQDT calculations. The positions of the f states were determined in zero-field measurements and are known to an accuracy of better than 1 MHz at $n \approx 30$ \cite{doran24a}, and their uncertainties do not cause significant uncertainties in the Stark shifts. The wavefunctions of the vibrationally excited ($v^+=1$) ion-core states extend over a larger range of internuclear distances $R$ than for $v^+=0$, for which accurate quantum defects are known \cite{osterwalder04a,sprecher14b}. For $\ell=0-2$ at $n \approx 34$, the uncertainties in the quantum defects of $v^+=1$ Rydberg states cause uncertainties on the order of several GHz in the predicted positions of the p and d Rydberg states. The s states do not play a role here because at low fields they lie so far away from the high-$\ell$ manifold that they do not have any influence on the Stark shifts. The uncertainties of the zero-field energies are primarily caused by the limited knowledge of the $R$ dependence of the quantum defects, and to a much smaller extent by the spin-induced interactions. Systematic deviations thus primarily originate from errors in the field-free positions of the p and d states and scale linearly with these errors and with the electric field strength. For the values of $n$ around 30 and $\mathcal{F}$ = 1 V/cm of interest, the errors are in the low-MHz range on the low-energy side of the Stark manifolds. They rapidly decrease with increasing $k$ value and become negligible on the high-energy side (see Ref. \cite{hoelsch22a}). 

The possible errors of the field-free energies of high-$\ell$ ($\ell \geq 4$) states result from the approximations used in the long-range interaction model (see Section \ref{subsec:longrange}), i.e., the truncation of the multipole expansion for the interaction between the Rydberg electron and the ion core, the neglect of nonadiabatic couplings and of the spin-spin and exchange interactions, as well as rovibrational channel interactions. These effects rapidly decrease with $\ell$. Consequently, we use the g state to determine an upper bound for the errors of the long-range interaction model. To this end, the energies of the g states are determined in MQDT calculations based on quantum defects obtained from a long-range model including higher-order corrections and nonadiabatic effects \cite{jungen89a, jungen90a, jungen25p}. To assess the effects of the exchange interaction, the singlet and triplet quantum defects are artificially separated by an offset which corresponds to the upper bound of 800 kHz estimated for the exchange interaction of the 30g state (Section \ref{subsec:longrange}). In spectral regions not significantly perturbed by channel interactions, the comparison between the energy levels obtained in these MQDT calculations and the long-range model implemented here reveals an agreement at the level of a fractional uncertainty in the g-state quantum defects of $\delta \mu_\textrm{g} / \mu_\textrm{g} = 1.5 \cdot 10^{-2}$  (corresponding to 3 MHz at $n\approx$ 30).

\begin{figure}
    \centering
     {\includegraphics[trim=13cm 2.5cm 15cm 3cm, clip=true, width=\linewidth]{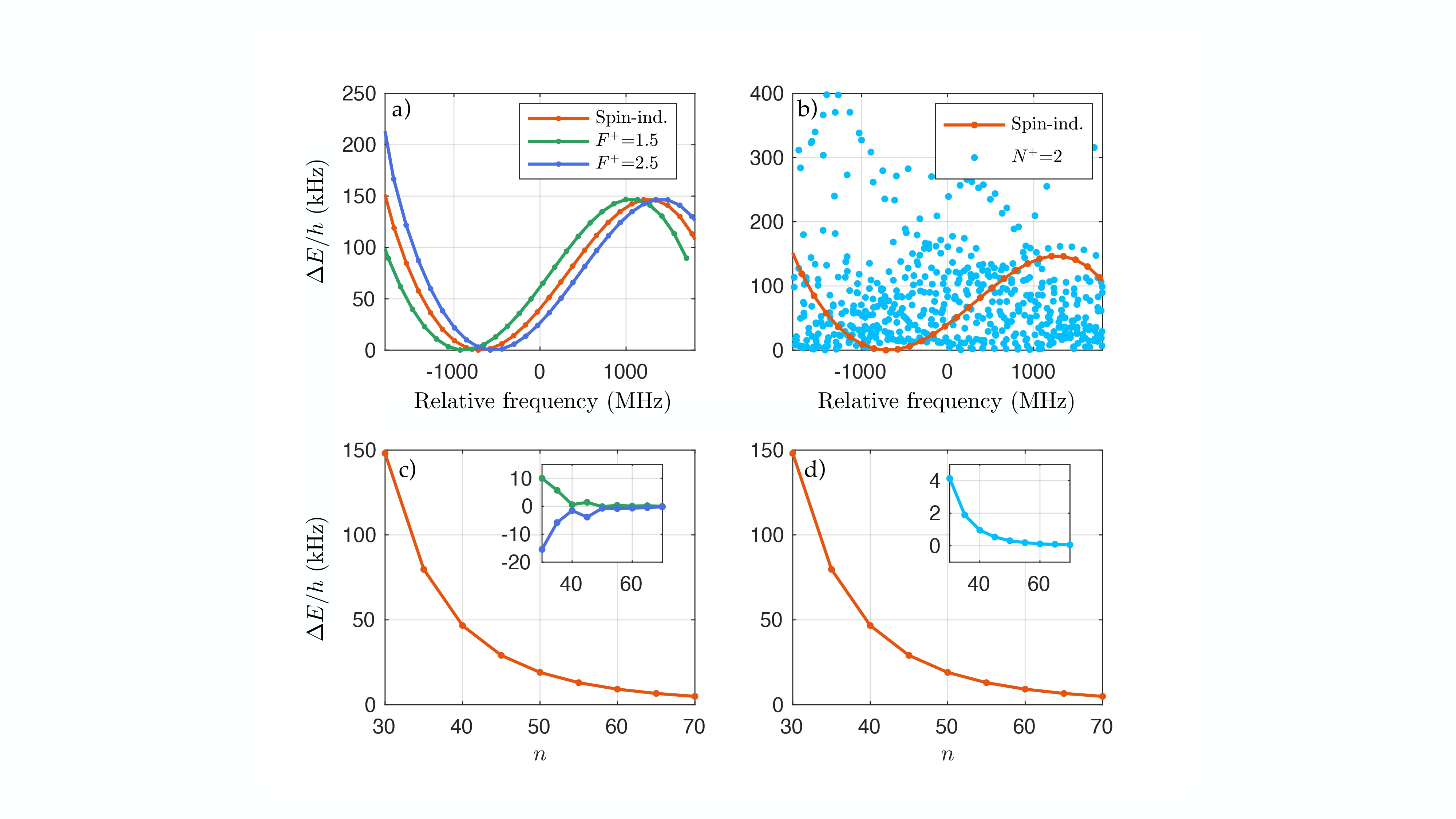}}
     \caption{Deviations ($\Delta E/h$) of calculated manifold positions resulting from a change of $\delta \mu_\textrm{g} / \mu_\textrm{g} = 1.5 \cdot 10^{-2}$ in the zero-field positions of g states at an electric field strength of $\mathcal{F}$ = 1 V/cm, for (a), (c): ortho-D$_2$, $v^+=1, N^+=0, I=2$ and (b),(d): para-H$_2$, $v^+=1, N^+=2, I=0$. The dots in the upper plots show the deviations of the different Stark states for $n=34$ for spin-free calculations (orange curves) and calculations including spins (a: green and dark-blue curves for $F^+=1.5$ and $F^+=2.5$, respectively, and b: bright blue data points). The horizontal axis gives positions of the Stark states for the reference $\mu_\textrm{g} $ values. The orange curves in the lower panels [(c), (d)] depict the evolution of the mean deviation of calculated positions as a function of $n$ for spin-free calculations. The insets illustrate the additional average deviations caused by the spins (same color coding as the upper plots).}
    \label{fig:sensi_g}
\end{figure}

Fig. \ref{fig:sensi_g} illustrates the changes $\Delta E/h$ in the computed positions of Rydberg-Stark states (for $\mathcal{F}$ = 1 V/cm) resulting from a change in the zero-field position of the 34g states by 3 MHz. Figs. \ref{fig:sensi_g}a and c (b and d) correspond to the cases of ortho-D$_2$, $v^+=1, N^+=0, I=2$ (para-H$_2$, $v^+=1, N^+=2, I=0$). 
The horizontal axis gives the Stark shifts for the reference $\mu_{\textrm{g}}$ value using the same relative-frequency scale as in Fig. \ref{fig:starkmaps} (\textit{y} axis). In both cases, the changes in Stark shifts are less than 400 kHz. For ortho-D$_2$, the effect on each class of $F^+$ states (green and dark-blue curves for $F^+=1.5$ and $F^+=2.5$, respectively) is the same as for the spin-independent ($I=0$) component (orange curve). The oscillating behavior reflects the evolution of the g character across the manifold. For para-H$_2$, the errors vary almost randomly from one Stark state to the next because the Rydberg-Stark states cannot be assigned to a given $F^+$ value, as explained above in the discussion of Fig. \ref{fig:starkmaps}. However, the deviations in the manifold positions obtained in a calculation for $N^+=2$ including the effects of spins (blue data points) are of the same order of magnitude as those obtained in the calculation neglecting spins (orange curve). The lower panels in Fig. \ref{fig:sensi_g} depict the evolution with $n$ of the average deviation of the Stark shifts across the entire manifold corresponding to a fractional uncertainty in the quantum defect of the g state of $\delta \mu_\textrm{g} / \mu_\textrm{g} = 1.5 \cdot 10^{-2}$, for ortho-D$_2$ (c) and para-H$_2$ (d). The orange curves show the results of spin-independent calculations and indicate that the deviations (at $\mathcal{F}$ = 1 V/cm) rapidly decrease with increasing $n$ value. The differences between the mean deviations obtained in computations including spins and these spin-independent calculations are shown in the insets, with the same color code as in the upper plots. These differences are in the low-kHz range at $n\approx 30$ and are negligible beyond $n=50$.

The errors resulting from the calculation of the electric-field-induced couplings are related to i) approximations regarding the basis states and ii) the calculation of wavefunctions corresponding to these basis states. Convergence on the kHz level in the calculation of manifold positions for typical values of $n$ ($\approx 30$) and $\mathcal{F}$ ($\approx$ 1 V/cm) is achieved when using all basis states ranging from $n-4$ to $n+4$. Because the experiments are performed for values of $n$ which are not in the vicinity of local perturbations caused by rovibrational channel interactions, all basis states (and therefore their radial wavefunctions) have, to a good approximation, the same vibrational ($v^+=1$) and rotational ($N^+=0$ for ortho-D$_2$ and $N^+=2$ for para-H$_2$) quantum numbers. Numerical evaluation of the wavefunctions of Rydberg states using the Numerov algorithm can result in divergence close to the ion core because of the strong mixing of states in the 
$\ket {(\ell s)j \left[ (S^+ N^+)J^+(I) \right] F^+F}$ basis set (see Fig. \ref{fig:basis_sets}). We truncate the wavefunctions either at the point of divergence, or at the polarizability radius of the H$_2^+$ ion \cite{hoelsch22a}. This truncation does not affect the calculated values of the Stark shifts by more than 20 kHz.

\section{Frame transformations between basis sets}
\label{sec:frametrafos} 

The basis sets used in each step of the calculations (see Fig. \ref{fig:basis_sets}) can be related using frame transformations consisting of sequential angular-momentum recoupling steps. The overall transformation between two basis sets is expressed as a product of frame transformations corresponding to each recoupling step. The quantum number $M_F$ is not indicated in these transformations, because it is irrelevant in the absence of an external field. As illustration, we derive here the transformation between the $\ket{(\ell s)j \left[ (S^+ N^+)J^+(I) \right] F^+F}$ and the $ \ket{(\ell N^+) N_{\Lambda} (sS^+)SJ)(I)F} $ basis sets (basis sets $\ket 3$ and $\ket 4$ in Fig. \ref{fig:basis_sets}). Fig. \ref{fig:trafo3} shows how the overall frame transformation can be decomposed into six successive steps in which, for simplicity, each intermediate angular-momentum coupling diagram is labeled only with the angular momenta that are recoupled in that step (shown in black). The first step (A) represents a transformation between Hund's cases (b) and (d) for $\Sigma$ ionic-state configurations \cite{jungen98a}, steps B and D-F are recouplings of angular momenta expressed by Wigner-6j symbols, and step C is a change of ordering in the coupling of two angular momenta. The equations corresponding to these steps are:

\begin{figure}
    \centering
     {\includegraphics[trim=0cm 37cm 16cm 1cm, clip=true, width=\linewidth]{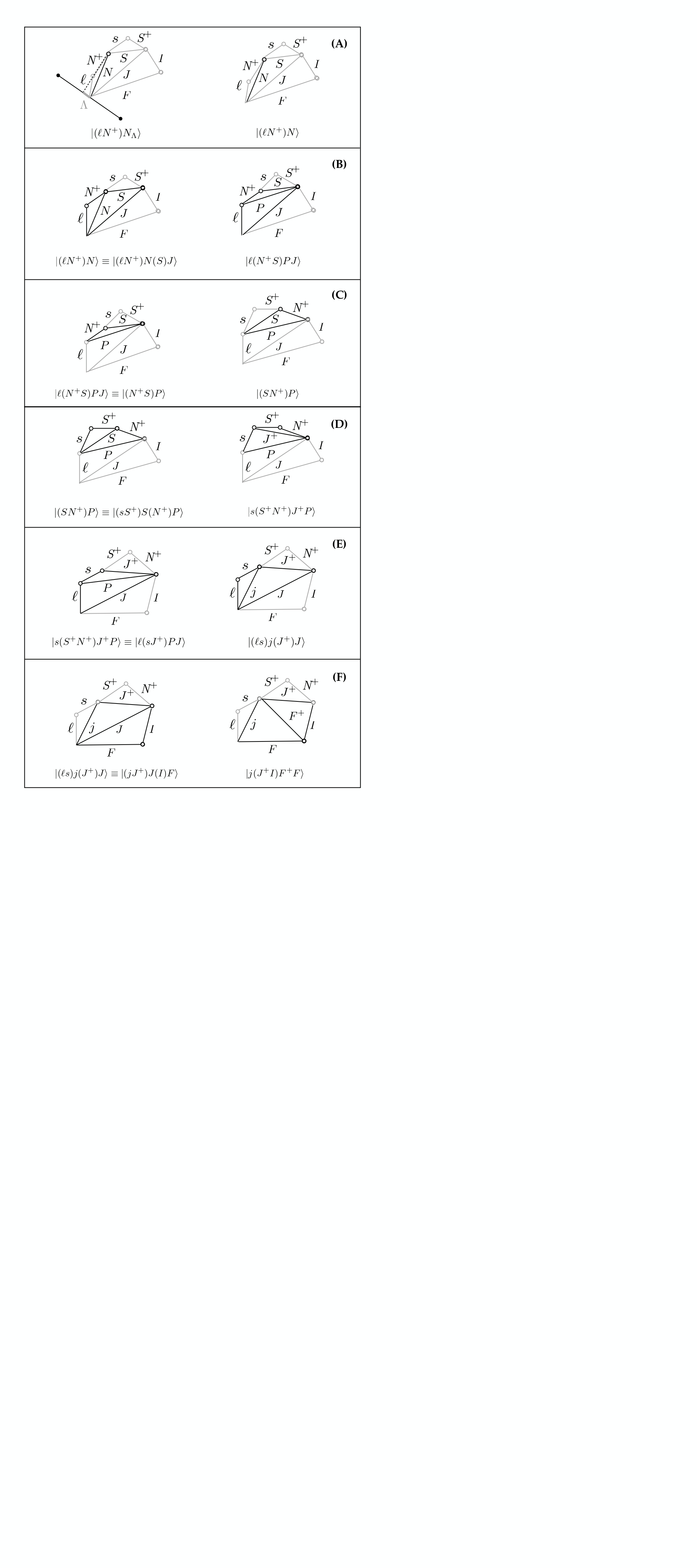}}
      \caption{Sequence of frame transformations between the $\ket{(\ell s)j (S^+ N^+)J^+(I)F^+F}$ and the $ \ket{(\ell N^+) N_{\Lambda} (sS^+)SJ(I)F} $ basis sets. For each intermediate step A-F, the initial and final basis states are labeled only with the angular momenta that are recoupled in that step (shown in black in the angular-momentum coupling schemes). The final basis state in each step becomes the initial basis state in the following one, and is relabeled every time. All other angular-momentum vectors are drawn in grey.}
    \label{fig:trafo3}
\end{figure}

\begin{equation*}
\braket{(\ell N^+)N_{\Lambda}|(\ell N^+)N} = (-1)^{\ell + \Lambda - N^+}
\end{equation*}
\begin{equation*}
\times \ThreeJ{\ell}{N}{N^+}{-\Lambda}{\Lambda}{0} \sqrt{2N^+ + 1} \sqrt{\frac{2}{1 + \delta_{\Lambda 0}}},
\label{eq:intens-stark-1}
\tag{A}
\end{equation*}
\begin{equation*}
\braket{(\ell N^+)N (S) J|\ell (N^+S)PJ} = \delta_{\ell \ell'} \delta_{N^+ N^{+}{'}} \delta_{SS'} \delta_{JJ'}
\end{equation*}
\begin{equation*}
\times (-1)^{\ell + N^+ + S + J} \sqrt{(2N+1)(2P+1)} \SixJ{\ell}{N^+}{N}{S}{J}{P},
\label{eq:intens-stark-2}
\tag{B}
\end{equation*}
\begin{equation*}
\braket{(N^+ S) P| (S N^+)P} = (-1) ^ {N^+ + S - P},
\label{eq:intens-stark-3}
\tag{C}
\end{equation*}
\begin{equation*}
\braket{(s S^+)S (N^+) P|s(S^+ N^+)J^+ P} = \delta_{ss'} \delta_{S^+ S^{+}{'}} \delta_{N^+N^{+}{'}} \delta_{PP'}
\end{equation*}
\begin{equation*}
\times (-1)^{s + S^+ + N^+ + P} \sqrt{(2S+1)(2J^++1)} \SixJ{s}{S^+}{S}{N^+}{P}{J^+},
\label{eq:intens-stark-4}
\tag{D}
\end{equation*}
\begin{equation*}
\braket{\ell (sJ^+)P J|(\ell s) j (J^+) J} = \braket{(\ell s) j (J^+) J|\ell (s  J^+) P J} 
\end{equation*}
\begin{equation*}
=\delta_{\ell \ell'} \delta_{ss'} \delta_{J^+ J^{+}{'}} \delta_{JJ'} 
\end{equation*}
\begin{equation*}
\times (-1) ^{\ell + s + J^+ + J } \sqrt{(2j+1)(2P+1)}\SixJ{\ell}{s}{j}{J^+}{J}{P},
\label{eq:intens-stark-5}
\tag{E}
\end{equation*}
and
\begin{equation*}
\braket{(jJ^+)J (I) F| j(J^+ I) F^+ F} = \delta_{jj'} \delta_{J^+ J^{+}{'}} \delta_{II'} \delta_{FF'} 
\end{equation*}
\begin{equation*}
\times (-1) ^{j + J^+ + I + F } \sqrt{(2J+1)(2F^++1)} \SixJ{j}{J^+}{J}{I}{F}{F^+}.
\label{eq:intens-stark-6}
\tag{F}
\end{equation*}

Equivalent transformations were performed between the other basis sets used in this work, and the corresponding expressions and diagrams are provided in the Appendix [Eqs. (IA-E) and (IIA-C), Figs. \ref{fig:trafo1} and \ref{fig:trafo2}].

\section{Conclusions}
In this work we have introduced a general method for the calculation of energy levels of molecular Rydberg states in static electric fields, including, for the first time, the effects of the fine and hyperfine structures. In order to separately assess the influence of a nonzero nuclear spin and a nonzero rotational angular momentum, we applied this procedure to calculate Rydberg-Stark manifolds of ortho-D$_2$ ($I=2, N^+=0, v^+=1$) and para-H$_2$ ($I=0, N^+=2, v^+=1$), respectively. 

The zero-field energy levels were calculated using multichannel-quantum-defect theory for the low-$\ell$ ($\ell \leq 3$) states and a long-range electrostatic model \cite{jungen69a, eyler83a} extended to account for the effects of electron and nuclear spins for $\ell \geq 4$ states. The effects of various zero-field interactions and their evolution with $\ell$ and $n$ could be assessed in this way, leading to the following conclusions: 
\\

i) For $\ell=0-2$ states, the energy-level structure is determined by the exchange interaction and, in the case of $N^+>0$, also by the electrostatic interaction between the molecular rotation and the orbital motion of the Rydberg electron;

ii) For $N^+=0$ and $\ell \geq 3$ states, the energy levels are to an excellent approximation separated by the hyperfine splittings of the ion core, and the total ion core spin $F^+ = G^+$ ($\vec{F}^+ = \vec{G}^+ = \vec{S}^+ + \vec{I}$) is a good quantum number (see Fig. \ref{fig:nevol}a); 

iii) For $N^+ > 0$ and $\ell \geq 3$, the energy-level pattern evolves as a function of $n$. At low-$n$ values ($n \approx 40$ for $\ell=5$ states), the electrostatic interaction  ($\vec{N} = \vec{\ell} + \vec{N}^+$) determines the level structure, and $N$ is a good quantum number. Because the $\vec{\ell} + \vec{N}^+$ electrostatic coupling scales as $n^{-3}$, the $n$-independent spin-rotation interaction in the ion core becomes dominant at high $n$ ($n >100$ for $\ell=5$ states) and $F^+ = J^+$ ($\vec{F}^+ =\vec{J}^+ = \vec{S}^+ + \vec{N}^+$) is the pattern-determining quantum number (see Fig. \ref{fig:nevol}b).

The electric-field-induced energy shifts were determined by diagonalizing the total Hamiltonian matrix $\hat{H} = \hat{H}_0  + e \mathcal{F} \hat{z}$, using a tensor-algebra formalism to evaluate the electric-field-induced couplings. Errors in these calculations can originate from a) uncertainties in the quantum defects and positions of the low-$\ell$ states from MQDT calculations, b) uncertainties in the positions of the high-$\ell$ states from a long-range polarization model, and c) approximations in the calculation of the eigenvalues of $\hat{H}$. An analysis of these sources of errors led to the following conclusions:
\\ 

a) The s levels are only weakly coupled to the linear Stark manifold and uncertainties in their positions do not affect the Stark shifts. The uncertainties in the zero-field positions of f states are very small and do not cause significant errors. The dominant errors originate from the uncertainties in the positions of the p and d levels. These errors primarily affect the positions of the Stark states with the smallest $k$ value and typical errors are below 100 kHz for $k>0$ Stark states at $n \geq 30$.

b) Uncertainties in the zero-field positions of the high-$\ell$ states rapidly decrease with increasing $\ell$ and cause maximal errors in the Stark shifts of less than 200 kHz at $n=34$ (see Fig. \ref{fig:sensi_g}).

c) Approximations in the radial wave functions of the Rydberg states and the neglect of off-diagonal matrix elements with $(n-n') \geq 4$ lead to errors below 20 kHz.

For comparison with experimental results and to clarify the role of the core angular momenta $I$ and $N^+$, we also presented a procedure to predict Rydberg-Stark spectra with the example of the resonant multiphoton excitation schemes $\text{X}\,^1\Sigma_g^+(v=0, N=0) \rightarrow \text{B}\,^1\Sigma_u^+(v=4,N=1) \rightarrow \text{GK}\,^1\Sigma_g^+\, (v=2,N=2) \rightarrow n \,\,[\text{D}_2^+\,\text{X}{^{+}}\,^{2} \Sigma_g^+ (v^+=1, N^+=0) \,]$ for ortho-D$_2$ and $\text{X}\,^1\Sigma_g^+(v=0, N=0) \rightarrow \text{B}\,^1\Sigma_u^+(v=4,N=1) \rightarrow \text{GK}\,^1\Sigma_g^+\, (v=2,N=0) \rightarrow n \,\,[\text{H}_2^+\,\text{X}{^{+}}\,^{2} \Sigma_g^+ (v^+=1, N^+=2) \,]$ for para-H$_2$. The spectral intensities were determined in the short-range coupling regime (basis set $\ket{4}$ in Fig. \ref{fig:basis_sets}) in which optical selection rules ($\Delta S = 0, \Delta \ell = \pm 1, \Delta F = 0, \pm1, \Delta \Lambda = 0, \pm1, \Delta M_F = 0, \pm1$) strongly restrict the available channels. Coherent and incoherent sums over the possible excitation pathways needed to be considered to account for interference effects and to enforce the conservation of $M_F$ in the homogeneous electric field, respectively (see Eqs. \eqref{eq:linestrength_gen} and \eqref{eq:threeph}). 

For the examples of H$_2$ and D$_2$ selected for this investigation, the calculations revealed a fundamental difference in the Stark effect in  $I=2, N^+=0$ (D$_2$) and $I=0, N^+=2$ (H$_2$) Rydberg states: in the absence of molecular rotation, the presence of a nonzero nuclear spin does not significantly influence the effect of the electric field on the Rydberg states. To an excellent approximation, the calculated linear Rydberg-Stark manifolds are independent and almost identical, regardless of the hyperfine state of the ion core. Consequently,  the hyperfine splitting of the ion core is the dominant pattern of the level structure, $F^+$ is a good quantum number in the presence of the electric field and the Stark states can be labeled by $k$. 

In contrast, when the ion core has a nonzero rotational angular momentum, the electrostatic coupling $\vec{N}^+ + \vec{\ell}$ of the molecular rotation with the orbital motion of the Rydberg electron competes with the spin-rotation interaction in the ion core and causes strong mixing between Rydberg-Stark states of different $J^+$ values. The spin-rotation interval of the ion core is no longer a dominant pattern of the level structure and $k$ is no longer a reliable label for the Stark states.

The approach presented in this article is expected to be generally applicable to the determination of rovibrational and hyperfine intervals in any diatomic molecule, independently of the existence of a nonzero permanent electric-dipole moment of the ion. The evaluation of possible sources of uncertainties in the calculations led to the conclusion that Stark states can be calculated with an accuracy of better than 200 kHz, which in turn implies that the ionic fine and hyperfine splittings can be determined from experimental Stark spectra at the same level of accuracy.

\begin{acknowledgments}
We thank Jean-Philippe Karr for useful advice on the theoretical treatment of the hyperfine structure of molecular hydrogen ions. This work is supported financially by the Swiss National Science Foundation (grant No.: 200021-236716) and the European Union’s Horizon Europe Research and Innovation Programme and the Participating States (funder ID: 10.13039/100019599, grant No.: 23FUN04 COMOMET).
\end{acknowledgments}

\appendix
\renewcommand{\thesection}{}
\section*{Appendix}
\begin{figure}
    \centering
     {\includegraphics[trim=0cm 44cm 16cm 1cm, clip=true, width=\linewidth]{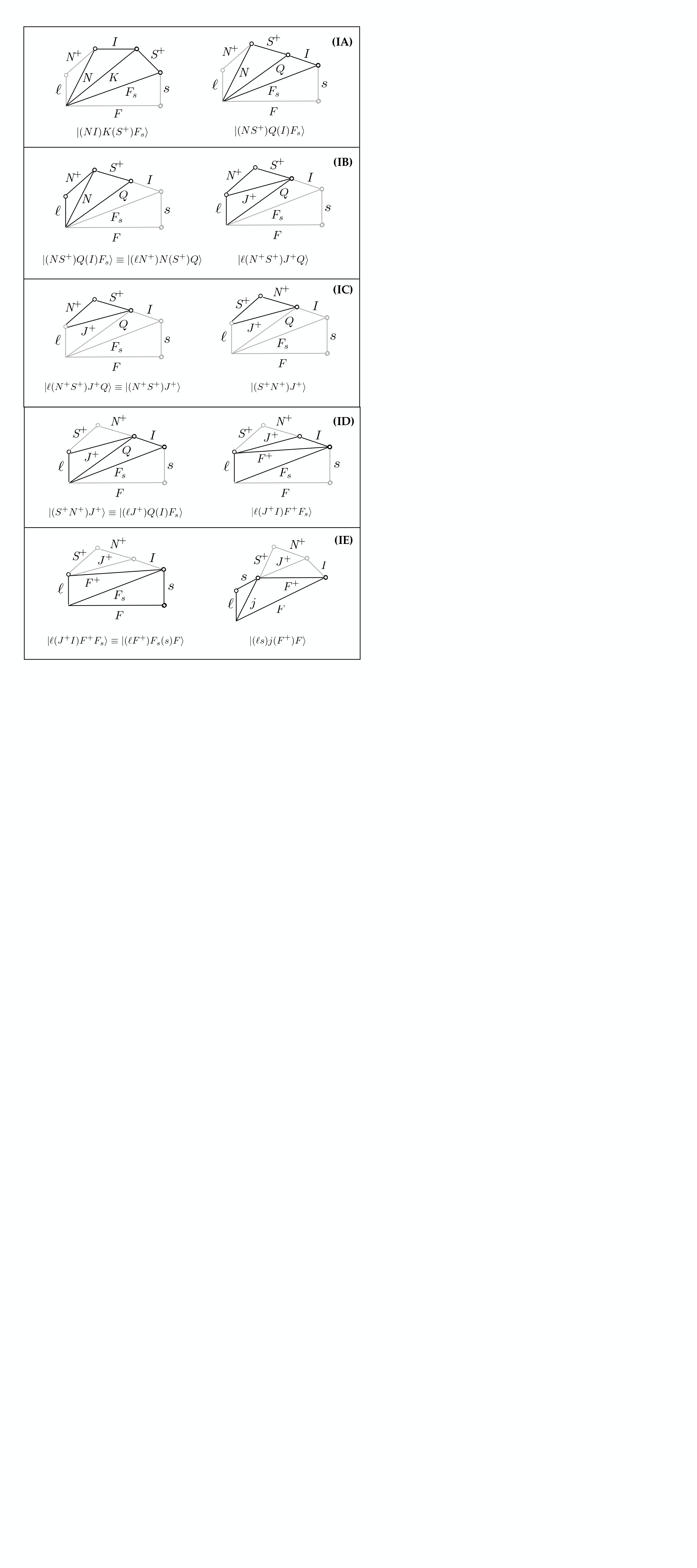}}
      \caption{Sequence of frame transformations between the $\ket{(\ell s)j (S^+ N^+)J^+(I)F^+F}$ and the $\ket{(\ell N^+)N(I)K(S^+)F_s(s)F}$ basis sets.}
    \label{fig:trafo1}
\end{figure}

\begin{figure}
    \centering
     {\includegraphics[trim=0cm 56cm 16cm 1cm, clip=true, width=\linewidth]{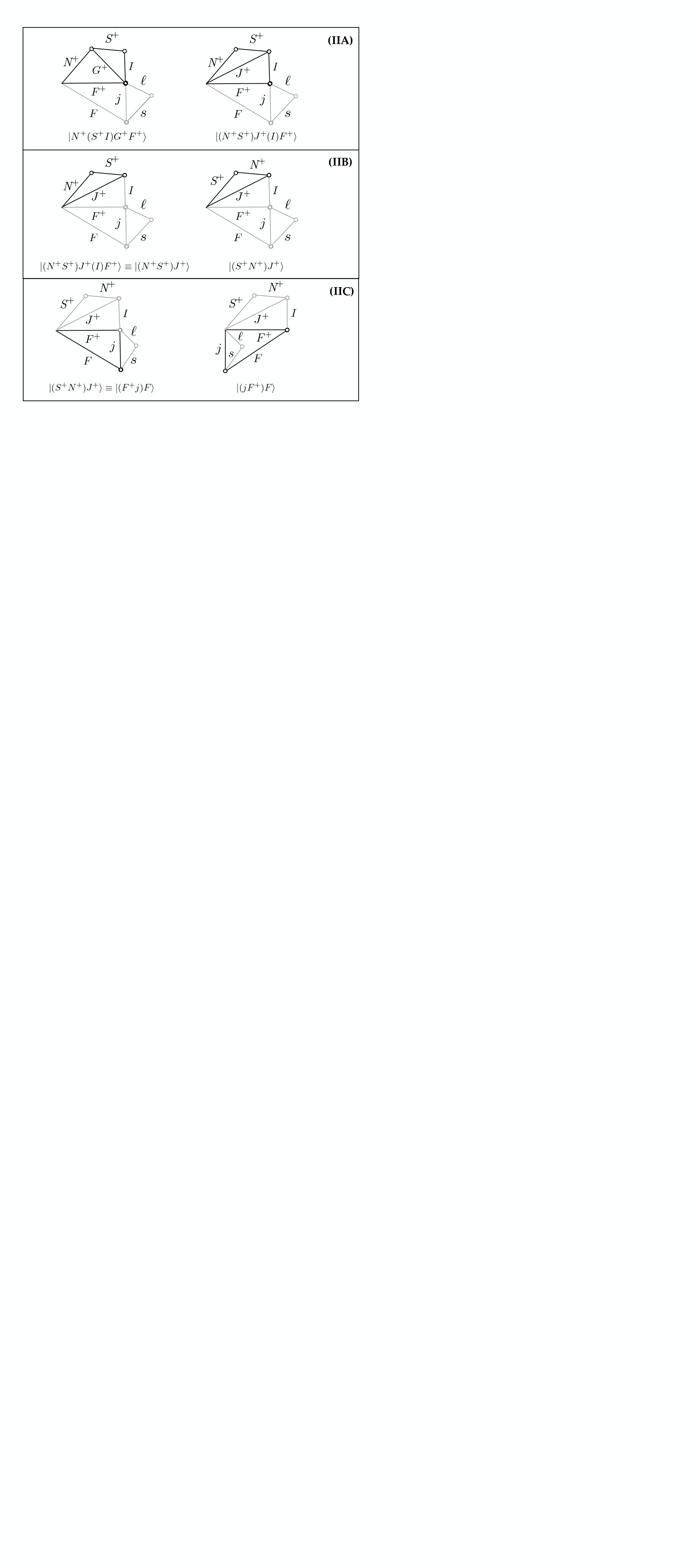}}
      \caption{Sequence of frame transformations between the $\ket{(\ell s)j (S^+ N^+)J^+(I)F^+F}$ and the $\ket {\left[N^+(S^+I)G^+ \right]F^+(\ell s)j F}$ basis sets.}
    \label{fig:trafo2}
\end{figure}

Eqs. (IA-E) and (IIA-C) and Figs. \ref{fig:trafo1} and \ref{fig:trafo2} give the expressions and corresponding diagrams of the frame transformations between the $\ket{(\ell s)j (S^+ N^+)J^+(I)F^+F}$ and the $\ket{(\ell N^+)N(I)K(S^+)F_s(s)F}$ basis sets, and the $\ket{(\ell s)j (S^+ N^+)J^+(I)F^+F}$ and the$\ket {\left[N^+(S^+I)G^+ \right]F^+(\ell s)j F}$ basis sets, respectively.

\begin{equation*}
\braket{(NI)K(S^+)F_s|(NS^+)Q(I)F_s} = \delta_{NN'}\delta_{II'}\delta_{S^+S^{+}{'}}\delta_{F_sF_s'}
\end{equation*}
\begin{equation*}
\times(-1)^{I+S^++K+Q}\sqrt{(2K+1)(2Q+1)} \SixJ{N}{I}{K}{F_s}{S^+}{Q}
\tag{IA}
\label{eq:polmol-stark-1}
\end{equation*}
\begin{equation*}
\braket{(\ell N^+)N(S^+)Q|\ell(N^+S^+)J^+Q} = \delta_{\ell \ell'}\delta_{N^+N^{+}{'}}\delta_{S^+S^{+}{'}}\delta_{QQ'}
\end{equation*}
\begin{equation*}
\times(-1)^{\ell+N^++S^++Q}\sqrt{(2N+1)(2J^++1)} \SixJ{\ell}{N^+}{N}{S^+}{Q}{J^+}
\label{eq:polmol-stark-2}
\tag{IB}
\end{equation*}
\begin{equation*}
\braket{(N^+S^+)J^+|(S^+N^+)J^+} = (-1)^{N^+ + S^+ - J^+}
\tag{IC}
\label{eq:polmol-stark-3}
\end{equation*}
\begin{equation*}
\braket{(\ell J^+)Q(I)F_s|\ell(J^+I)F^+F_s} = \delta_{\ell \ell'}\delta_{J^+J^{+}{'}}\delta_{II'}\delta_{F_s F_s'}
\end{equation*}
\begin{equation*}
\times(-1)^{\ell+J^++I+F_s}\sqrt{(2Q+1)(2F^++1)} \SixJ{\ell}{J^+}{Q}{I}{F_s}{F^+}
\tag{ID}
\label{eq:polmol-stark-4}
\end{equation*}
\begin{equation*}
\braket{(\ell F^+)F_s(s)F|(\ell s)j(F^+)F} = \delta_{\ell \ell'}\delta_{F^+F^{+}{'}}\delta_{ss}\delta_{F F'}
\end{equation*}
\begin{equation*}
\times(-1)^{F^+ + s + F_s + j}\sqrt{(2F_s+1)(2j+1)} \SixJ{\ell}{F^+}{F_s}{F}{s}{j}
\label{eq:polmol-stark-5}
\tag{IE}
\end{equation*}
\hrulefill

\begin{equation*}
\braket{N^+ (S^+ I) G^+ F^+ | (N^+ S^+) J^+ (I) F} = \delta_{N^+ N^{+}{'}}\delta_{S^+S^{+}{'}}\delta_{II'}\delta_{F^+ F^{+}{'}}
\end{equation*}
\begin{equation*}
\times (-1)^{N^+ + S^+  +I + F^+}\sqrt{(2J^++1)(2G^++1)} \SixJ{N^+}{S^+}{J^+}{I}{F^+}{G^+}
\label{eq:mqdt-stark-1}
\tag{IIA}
\end{equation*}
\begin{equation*}
\braket{(N^+ S^+) J^+ | (S^+ N^+) J^+} = (-1) ^{N^+ + S^+ - J^+} 
\label{eq:mqdt-stark-2}
\tag{IIB}
\end{equation*}
\begin{equation*}
\braket{(F^+ j) F | (j F^+) F} = (-1) ^{F^+ + j - F} 
\label{eq:mqdt-stark-3}
\tag{IIC}
\end{equation*}

\clearpage

%

\clearpage

\begin{figure*}
\renewcommand{\thefigure}{S1} 
    \centering
     {\includegraphics[trim=4.5cm 0.4cm 2cm 0cm, clip=true, angle=90, width=0.8\linewidth]{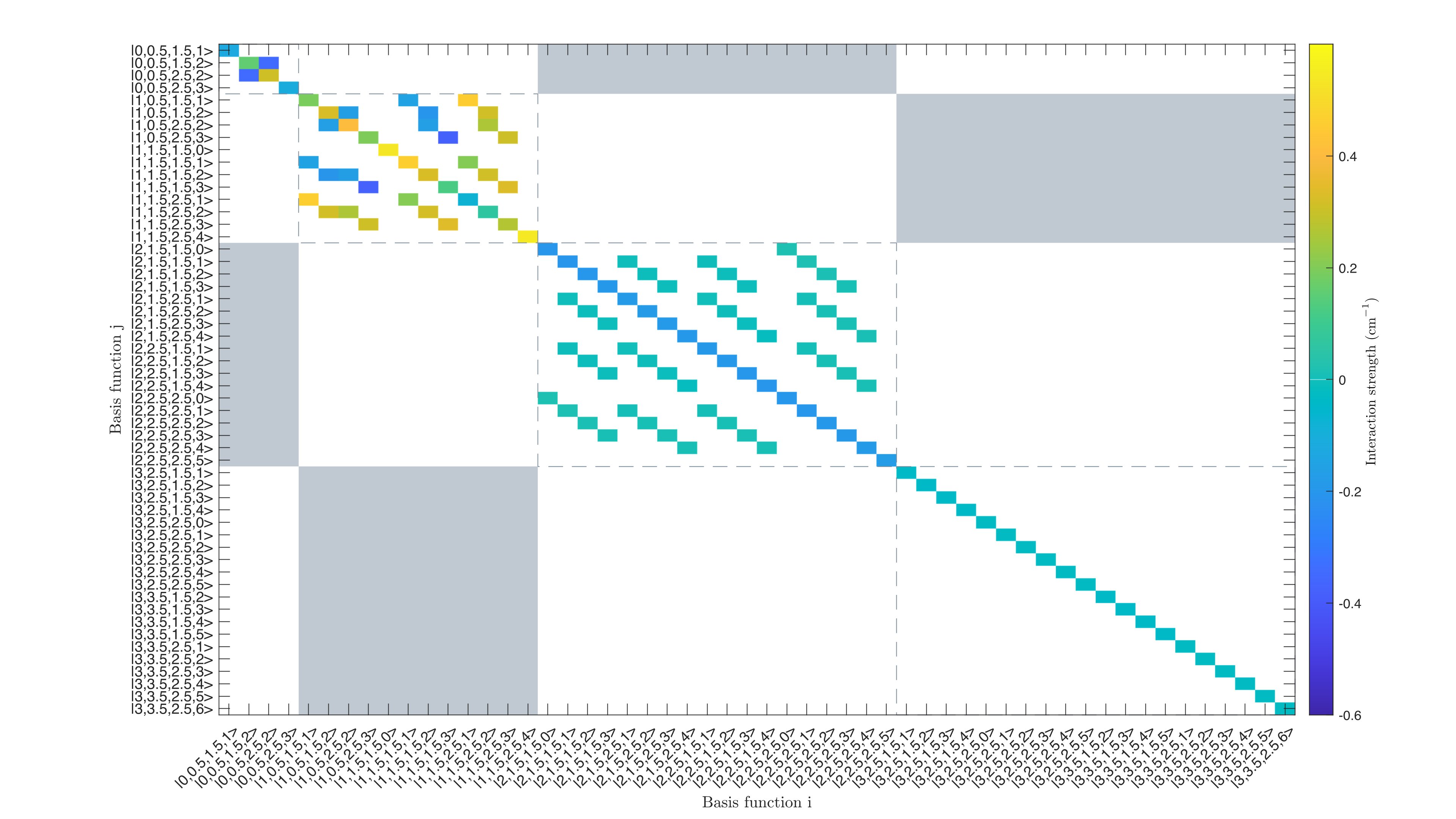}}
      \caption{$\ell=0-3, n = 34, v^+=1$ zero-field matrix for ortho-D$_2$, $I=2, N^+=0$ obtained from MQDT calculations including spins. The blocks are labeled by $\ket{\ell jF^+F}$.}
    \label{fig:matrix_Hzero_big_od2}
\end{figure*}

\begin{figure*}
\renewcommand{\thefigure}{S2} 
    \centering
     {\includegraphics[trim=4.5cm 0.4cm 2cm 0cm, clip=true, angle=90, width=0.8\linewidth]{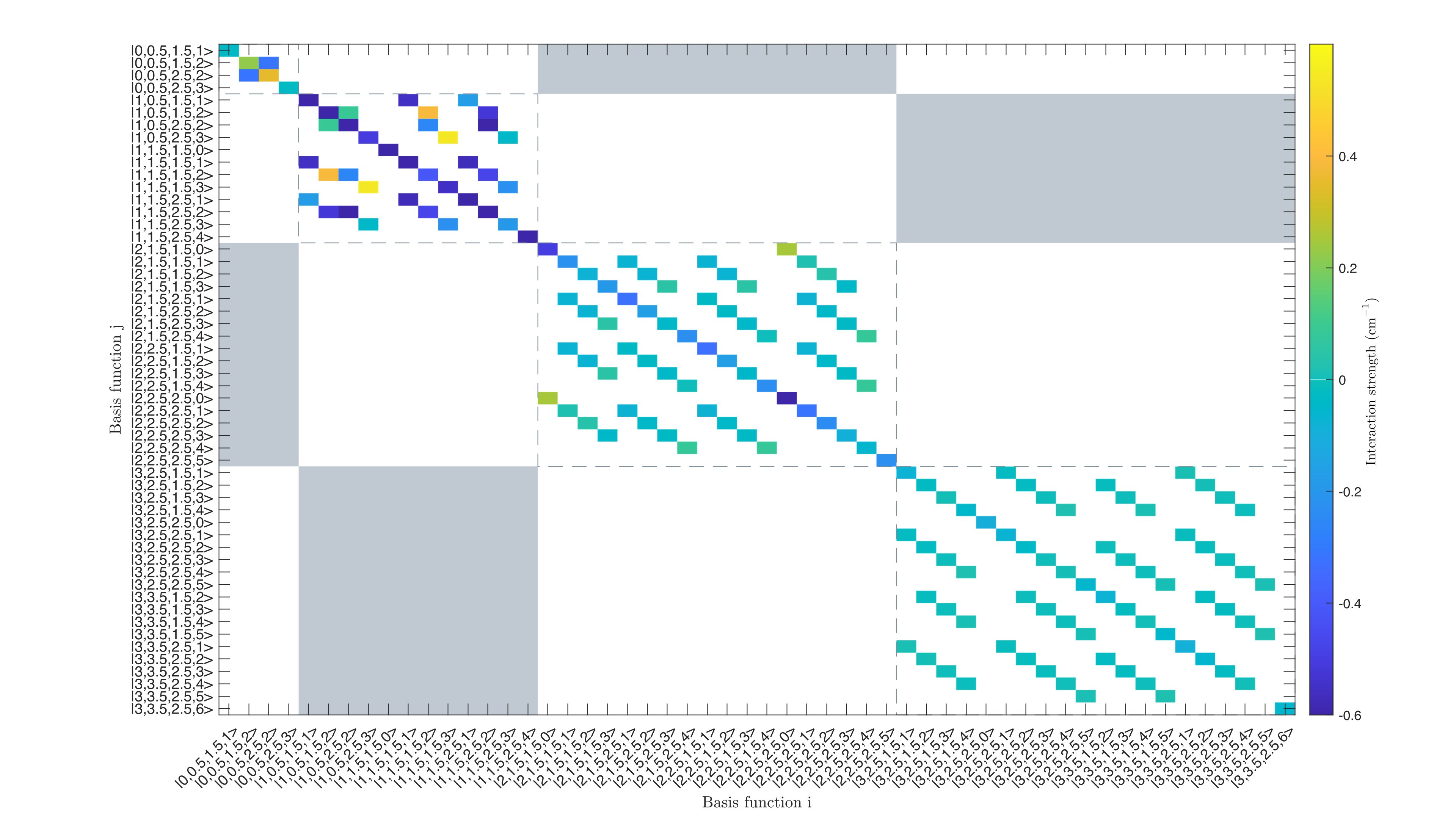}}
      \caption{$\ell=0-3, n = 34, v^+=1$ zero-field matrix for para-H$_2$, $I=0, N^+=2$ obtained from MQDT calculations including spins. The blocks are labeled by $\ket{\ell jF^+F}$.}
    \label{fig:matrix_Hzero_big_ph2}
\end{figure*}

\begin{figure*}
\renewcommand{\thefigure}{S3} 
    \centering
     {\includegraphics[trim=4.5cm 0.4cm 2cm 0cm, clip=true, angle=90, width=0.8\linewidth]{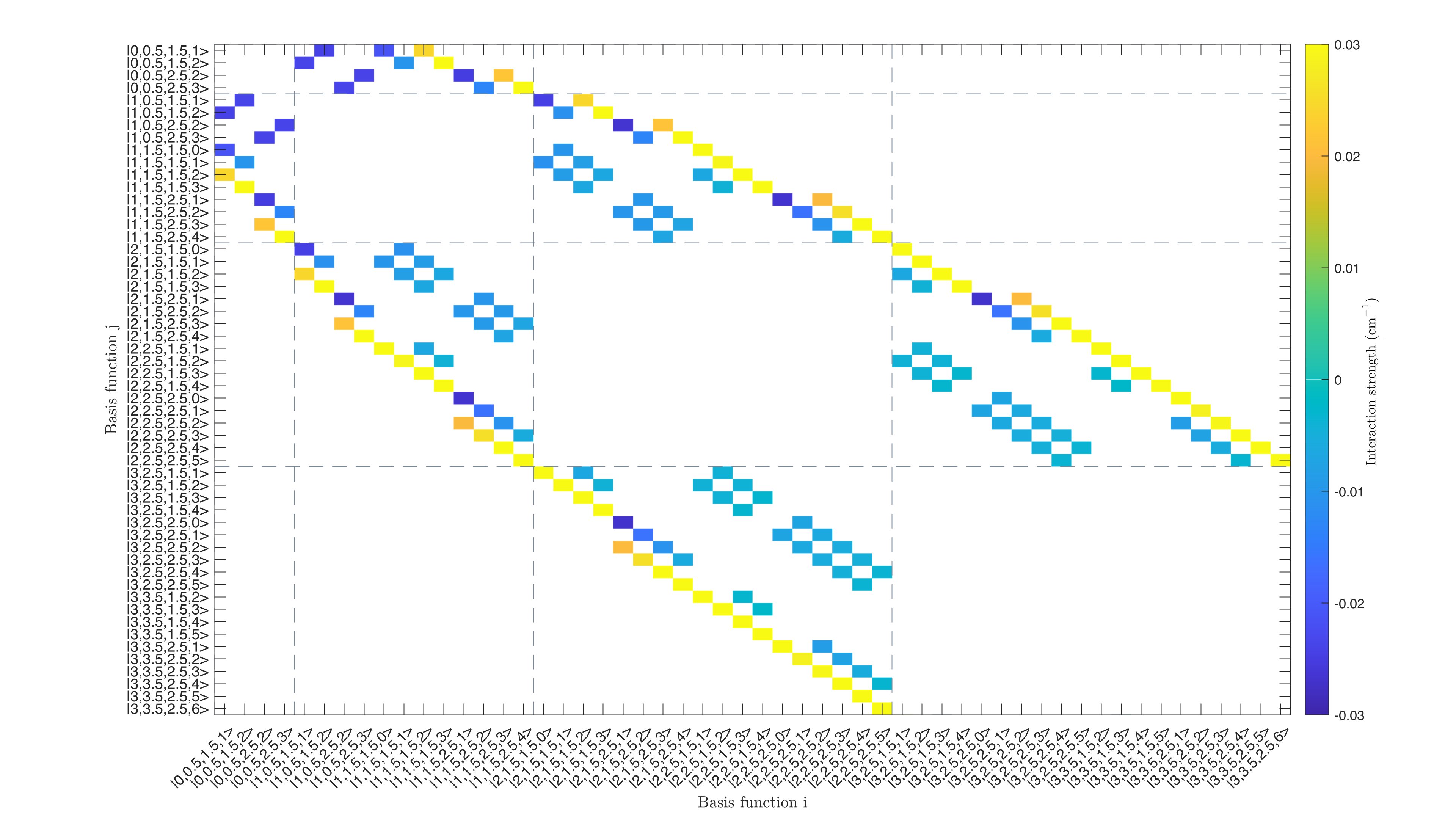}}
      \caption{$\ell=0-3, n = 34, v^+=1$ Stark matrix for para-H$_2$, $I=0, N^+=2$, $n = 34, v^+=1, M_F=0$, and an electric field strength of 1 V/cm. The blocks are labeled by $\ket{\ell jF^+F}$.}
    \label{fig:matrix_Hstark_big_ph2}
\end{figure*}

\begin{figure*}
\renewcommand{\thefigure}{S4} 
    \centering
     {\includegraphics[trim=4.5cm 0.4cm 2cm 0cm, clip=true, angle=90, width=0.8\linewidth]{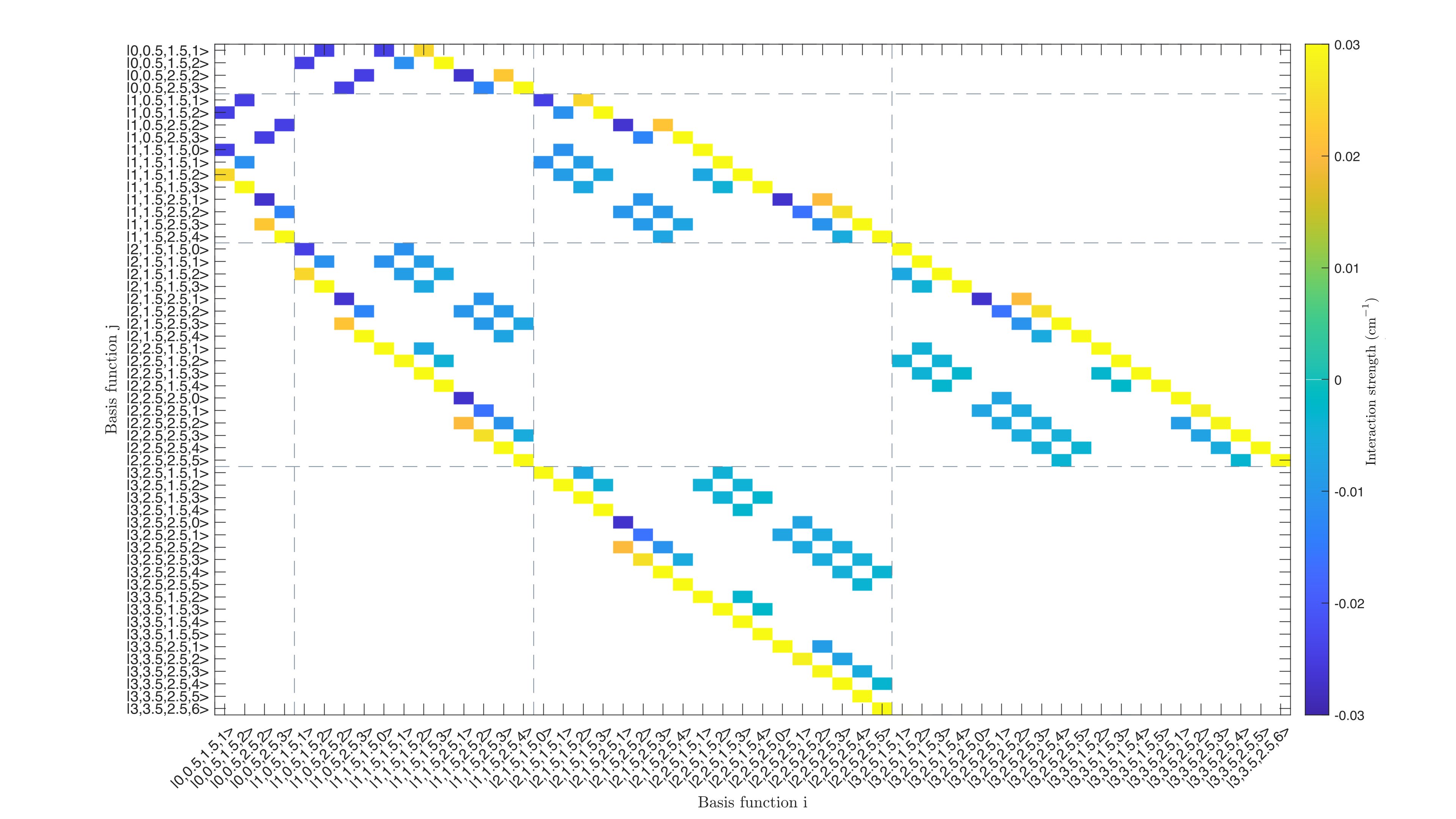}}
      \caption{$\ell=0-3, n = 34, v^+=1$ Stark matrix for ortho-D$_2$, $I=2, N^+=0$, $n = 34, v^+=1, M_F=0$, and an electric field strength of 1 V/cm. The blocks are labeled by $\ket{\ell jF^+F}$.}
    \label{fig:matrix_Hstark_big_od2}
\end{figure*}

\begin{widetext}
\section*{Supplemental Material}
This Supplemental material provides step-by-step derivations of: SI) the spin-rotation coupling term $c_e (\bm{N^+} \cdot \bm{S^+})$ evaluated in the $\ket{n v^+} \ket{(\ell N^+)N(I)K(S^+)F_s(s)FM_F}$ basis set [Eq. (9) in the main text]; SII) the hyperfine coupling term $b_F (\bm{I} \cdot \bm{S^+})$ evaluated in the $\ket{n v^+} \ket{(\ell N^+)N(I)K(S^+)F_s(s)FM_F}$ basis set [Eq. (10) in the main text], SIII-IV) the spin-orbit $\hat{H}_{\textrm{so}}$ and the other-spin$-$orbit $\hat{H}_{\textrm{oso}}$ interactions evaluated in the $\ket{n v^+} \ket{(\ell N^+)N(I)K(S^+)F_s(s)FM_F}$ basis set [Eq. (12) in the main text], as well as SV) the elements of $D_{-pq}^{(1)}  (\omega)^*$ evaluated in the $ \ket {(\ell N^+)N_\Lambda (sS^+)S J (I) F M_F}$ basis set [Eq. (17) in the main text]. Figs. \ref{fig:matrix_Hzero_big_od2}, \ref{fig:matrix_Hzero_big_ph2}, \ref{fig:matrix_Hstark_big_ph2} show enlarged versions of Figs. 2 and 7 of the article, where the complete labels of the basis states with the values  of $j, F^+$ and $F$ are given. Fig. \ref{fig:matrix_Hstark_big_od2} shows the elements of the Stark matrix for ortho-D$_2$, $\ell=0-3, n = 34, v^+=1$, $I=2, N^+=0$, $M_F=0$ Rydberg states at an electric field strength of 1 V/cm.

\begin{equation*}
c_e \braket{(\ell N^+)N(I)K(S^+)F_s(s)FM_F| \bm{T^1}(N^+) \cdot \bm{T^1}(S^+)|(\ell ' N^{+}{'})N'(I')K'(S^{+}){'}F_s'(s')F'M_F'} 
\end{equation*}
\begin{equation*}
=c_e  \delta_{M_F M_F'}\delta_{FF'} \delta_{F_sF_s'} (-1)^{K' + S^++F_s} \SixJ{K}{S^+}{F_s}{S^{+}{'}}{K'}{1}  
\end{equation*}
\begin{equation*}
\times \braket{(\ell N^+)N(I)K||\bm{T^1}(N^+)||(\ell ' N^{+}{'})N'(I')K'}  \braket{S^+||\bm{T^1}(S^+)||S^{+}{'}} 
\end{equation*}
\begin{equation*}
=c_e  \delta_{M_F M_F'}\delta_{FF'} \delta_{F_sF_s'} \delta_{I I'} (-1)^{K' + S^++F_s +N + I + K +1 } \SixJ{K}{S^+}{F_s}{S^{+}{'}}{K'}{1}  \SixJ{N}{K}{I}{K'}{N{'}}{1} 
\end{equation*}
\begin{equation*}
\times \sqrt{(2K+1)(2K'+1)} \braket{(\ell N^+) N ||\bm{T^1}(N^+)||(\ell'N^{+}{'})N'} \braket{S^+||\bm{T^1}(S^+)||S^{+}{'}}  
\end{equation*}
\begin{equation*}
=c_e  \delta_{M_F M_F'}\delta_{FF'} \delta_{F_sF_s'} \delta_{I I'} \delta_{\ell \ell'} (-1)^{K' + S^++F_s +N + I + K +1 + \ell + N^{+}{'} + N+1} \SixJ{K}{S^+}{F_s}{S^{+}{'}}{K'}{1}  \SixJ{N}{K}{I}{K'}{N{'}}{1} \SixJ{N^+}{N}{\ell}{N'}{N^{+}{'}}{1} 
\end{equation*}
\begin{equation*}
\times \sqrt{(2K+1)(2K'+1)(2N+1)(2N'+1)} \braket{N^+ ||\bm{T^1}(N^+)||N^{+}{'}} \braket{S^+||\bm{T^1}(S^+)||S^{+}{'}}  
\end{equation*}
\begin{equation*}
=c_e  \delta_{M_F M_F'}\delta_{FF'} \delta_{F_sF_s'} \delta_{I I'} \delta_{\ell \ell'} \delta_{N^+ N^{+}{'}} \delta_{S^+ S^{+}{'}} (-1)^{K' + S^+ +F_s  + I + K +  \ell + N^{+}{'}} \SixJ{K}{S^+}{F_s}{S^+}{K'}{1} \SixJ{N}{K}{I}{K'}{N{'}}{1} \SixJ{N^+}{N}{\ell}{N'}{N^{+}{'}}{1} 
\end{equation*}
\begin{equation}
\times \sqrt{(2K+1)(2K'+1)(2N+1)(2N'+1)(2N^++1)N^+(N^++1)(2S^++1)S^+(S^++1)}.
\label{eq:Hhfs9_ext}
\tag{SI}
\end{equation}

\hrulefill

\begin{equation*}
b_F\braket{(\ell N^+)N(I)K(S^+)F_s(s)FM_F| \bm{T^1}(I) \cdot \bm{T^1}(S^+)|(\ell ' N^{+}{'})N'(I')K'(S^{+}){'}F_s'(s')F'M_F'} 
\end{equation*}
\begin{equation*}
=b_F \delta_{M_F M_F'}\delta_{FF'} \delta_{F_sF_s'} (-1)^{K' + S^++F_s} \SixJ{K}{S^+}{F_s}{S^{+}{'}}{K'}{1} 
\end{equation*}
\begin{equation*}
 \times \braket{(\ell N^+)N(I)K||\bm{T^1}(I)||(\ell ' N^{+}{'})N'(I')K'}  \braket{S^+||\bm{T^1}(S^+)||S^{+}{'}} 
\end{equation*}
\begin{equation*}
=b_F \delta_{M_F M_F'}\delta_{FF'} \delta_{F_sF_s'} \delta_{NN'} (-1)^{K' + S^++F_s +N + I' + K +1 } \SixJ{K}{S^+}{F_s}{S^{+}{'}}{K'}{1}  \SixJ{I}{K}{N}{K'}{I{'}}{1} 
\end{equation*}
\begin{equation*}
\times \sqrt{(2K+1)(2K'+1)} \braket{I||\bm{T^1}(I)| I '} \braket{S^+||\bm{T^1}(S^+)||S^{+}{'}}  
\end{equation*}
\begin{equation*}
=b_F \delta_{M_F M_F'}\delta_{FF'} \delta_{F_sF_s'} \delta_{NN'}  \delta_{II'} \delta_{S^+ S^{+}{'}}  (-1)^{K' + S^++F_s +N + I' + K +1 }  \SixJ{K}{S^+}{F_s}{S^{+}{'}}{K'}{1}  \SixJ{I}{K}{N}{K'}{I{'}}{1}
\end{equation*}
\begin{equation}
\times \sqrt{(2K+1)(2K'+1)(2I+1)I(I+1)(2S^++1)S^+(S^++1)}.
\label{eq:Hhfs10}
\tag{SII}
\end{equation}

\hrulefill

\begin{equation*}
 \braket{(\ell N^+)N(I)K(S^+)F_s(s)FM_F|\bm{T^1}(\ell) \cdot \bm{T^1}(s)|(\ell ' N^{+}{'})N'(I')K'(S^{+}){'}F_s'(s')F'M_F'} 
\label{eq:Hhso4}
\end{equation*}

\begin{equation*}
=\delta_{M_F M_F'}\delta_{FF'} (-1)^{F_s' + s + F} \SixJ{F_s}{s}{F}{s'}{F_s'}{1} 
\label{eq:Hso5}
\end{equation*}

\begin{equation*}
\times \braket{(\ell N^+)N(I)K(S^+)F_s||\bm{T^1}(\ell)||(\ell ' N^{+}{'})N'(I')K'(S^{+}{'})F_s'}  \braket{s||\bm{T^1}(s)||s'}
\label{eq:Hso6}
\end{equation*}

\begin{equation*}
=\delta_{M_F M_F'}\delta_{FF'}\delta_{S^+S^{+}{'}} (-1)^{F_s' + s + F + K + S^+ + F_s + 1}\SixJ{F_s}{s}{F}{s'}{F_s'}{1}  \SixJ{K}{F_s}{S^+}{F_s'}{K'}{1}
\label{eq:Hso7}
\end{equation*}

\begin{equation*}
\times  \sqrt{(2F_s+1)(2F_s'+1)}\braket{(\ell N^+)N(I)K||\bm{T^1}(\ell)||(\ell ' N^{+}{'})N'(I')K'}  \braket{s||\bm{T^1}(s)||s'}
\label{eq:Hso8}
\end{equation*}

\begin{equation*}
=\delta_{M_F M_F'}\delta_{FF'}\delta_{S^+S^{+}{'}} \delta_{II'}(-1)^{F_s' + s + F+ K + S^+ + F_s + 1 + N + I + K' + 1}\SixJ{F_s}{s}{F}{s'}{F_s'}{1}  \SixJ{K}{F_s}{S^+}{F_s'}{K'}{1} \SixJ{N}{K}{I}{K'}{N'}{1} 
\label{eq:Hso81}
\end{equation*}

\begin{equation*}
 \times \sqrt{(2F_s+1)(2F_s'+1)(2K+1)(2K'+1)}\braket{(\ell N^+)N||\bm{T^1}(\ell)||(\ell ' N^{+}{'})N'}  \braket{s||\bm{T^1}(s)||s'}
\label{eq:Hso182}
\end{equation*}

\begin{equation*}
=\delta_{M_F M_F'}\delta_{FF'}\delta_{S^+S^{+}{'}} \delta_{II'} \delta_{N^+N^{+}{'}} \delta_{\ell \ell'} \delta_{ss'}(-1)^{F_s' + s + F+ K + S^+ + F_s + N + I + K'  + \ell + N^+ + N'+1}  
\label{eq:Hso9}
\end{equation*}

\begin{equation*}
\times \SixJ{F_s}{s}{F}{s'}{F_s'}{1}  \SixJ{K}{F_s}{S^+}{F_s'}{K'}{1} \SixJ{N}{K}{I}{K'}{N'}{1}  \SixJ{\ell}{N}{N^+}{N'}{\ell'}{1} 
\label{eq:Hso10}
\end{equation*}

\begin{equation}
 \times \sqrt{(2F_s+1)(2F_s'+1)(2K+1)(2K'+1)(2N+1)(2N'+1)(2\ell+1)\ell(\ell+1)(2s+1)s(s+1)}.
\label{eq:Hso11}
\tag{SIII}
\end{equation}

\hrulefill

\begin{equation*}
\braket{(\ell N^+)N(I)K(S^+)F_s(s)FM_F|\bm{T^1}(\ell) \cdot \bm{T^1}(S^+)|(\ell ' N^{+}{'})N'(I')K'(S^{+}){'}F_s'(s')F'M_F'} 
\label{eq:Hso15}
\end{equation*}

\begin{equation*}
= \delta_{M_F M_F'}\delta_{FF'} \delta_{F_sF_s'} (-1)^{K' + S^++F_s} \SixJ{K}{S^+}{F_s}{S^{+}{'}}{K'}{1}  
\label{eq:Hso16}
\end{equation*}

\begin{equation*}
\times \braket{(\ell N^+)N(I)K||\bm{T^1}(\ell)||(\ell ' N^{+}{'})N'(I')K'}  \braket{S^+||\bm{T^1}(S^+)||S^{+}{'}} 
\label{eq:Hso17}
\end{equation*}

\begin{equation*}
=\delta_{M_F M_F'}\delta_{FF'} \delta_{F_sF_s'} \delta_{II'} (-1)^{K' + S^++F_s + N + I + K' + 1} \SixJ{K}{S^+}{F_s}{S^{+}{'}}{K'}{1}  \SixJ{N}{K}{I}{K'}{N'}{1} 
\label{eq:Hso171}
\end{equation*}

\begin{equation*}
\times \sqrt{(2K+1)(2K'+1)} \braket{(\ell N^+) N ||\bm{T^1}(\ell)|| (\ell' N^{+}{'}) N'} \braket{S^+||\bm{T^1}(S^+)||S^{+}{'}}  
\label{eq:Hso172}
\end{equation*}

\begin{equation*}
=\delta_{M_F M_F'}\delta_{FF'} \delta_{F_sF_s'} \delta_{II'} \delta_{N^+ N^{+}{'}} (-1)^{K' + S^++F_s +  N + I + K' + 1 + \ell + N^+ + N' + 1} \SixJ{K}{S^+}{F_s}{S^{+}{'}}{K'}{1}   \SixJ{N}{K}{I}{K'}{N'}{1} \SixJ{\ell}{N}{N^+}{N'}{\ell'}{1} 
\label{eq:Hso18}
\end{equation*}

\begin{equation*}
\times \sqrt{(2K+1)(2K'+1)(2N+1)(2N'+1)} \braket{\ell ||\bm{T^1}(\ell)||\ell'} \braket{S^+||\bm{T^1}(S^+)||S^{+}{'}}  
\label{eq:Hso19}
\end{equation*}

\begin{equation*}
=\delta_{M_F M_F'}\delta_{FF'} \delta_{F_sF_s'} \delta_{II'} \delta_{N^+ N^{+}{'}}  \delta_{\ell \ell'} \delta_{S^+S^{+}{'}}(-1)^{K' + S^++F_s +  N + I + K' + \ell + N^+ + N' } \SixJ{K}{S^+}{F_s}{S^{+}{'}}{K'}{1} \SixJ{N}{K}{I}{K'}{N'}{1}  \SixJ{\ell}{N}{N^+}{N'}{\ell'}{1} 
\label{eq:Hso20}
\end{equation*}

\begin{equation}
\times \sqrt{(2K+1)(2K'+1)(2N+1)(2N'+1)(2S^++1)S^+(S^++1)(2\ell+1)\ell(\ell+1)}.
\label{eq:Hso21}
\tag{SIV}
\end{equation}

\hrulefill

\begin{equation*}
\braket{ (N_{\Lambda}S) J (I) F {M_F}|D_{-pq}^{(1)}  (\omega)^* | (N'_{\Lambda'}S') J' (I') F' {M_F}'}  
\end{equation*}
\begin{equation*}
=(-1)^{F-M_F}\ThreeJ{F}{1}{F'}{-M_F}{-p}{M_{F}'} \braket{ (N_{\Lambda}S) J (I) F  || D_{q}^{(1)}  (\omega)^* ||  (N'_{\Lambda'}S') J' (I') F'}
\end{equation*}
\begin{equation*}
= \delta_{I I'} (-1)^{F - M_F + J + I + F' + 1} \ThreeJ{F}{1}{F'}{-M_F}{-p}{M_{F}'} \SixJ{J}{F}{I}{F'}{J'}{1} \sqrt{(2F+1)(2F'+1)} \braket{ (N_{\Lambda}S) J  || D_{q}^{(1)}  (\omega)^* ||  (N'_{\Lambda'}S') J'}
\end{equation*}
\begin{equation*}
=\delta_{I I'} \delta_{SS'} (-1)^{F - M_F +  J + I + F'  + N + S + J' } \ThreeJ{F}{1}{F'}{-M_F}{-p}{M_{F}'} \SixJ{J}{F}{I}{F'}{J'}{1}  \SixJ{N}{J}{S}{J'}{N'}{1}  
\end{equation*}
\begin{equation*}
\times \sqrt{(2F+1)(2F'+1)(2J+1)(2J'+1)} \braket{ N_{\Lambda} || D_{q}^{(1)}  (\omega)^* ||  N'_{\Lambda'}}
\end{equation*}
\begin{equation*}
= \delta_{I I'} \delta_{SS'} (-1)^{F - M_F + J + I + F' + N + S + J'  + N - \Lambda} \ThreeJ{F}{1}{F'}{-M_F}{-p}{M_{F}'}  \SixJ{J}{F}{I}{F'}{J'}{1}  \SixJ{N}{J}{S}{J'}{N'}{1} \ThreeJ{N}{1}{N'}{-\Lambda}{q}{\Lambda}  
\label{eq:matel_seven}
\end{equation*}
\begin{equation}
\times \sqrt{(2F+1)(2F'+1)(2J+1)(2J'+1)(2N+1)(2N'+1)}.
\label{eq:int_full}
\tag{SV}
\end{equation}
\end{widetext}

\end{document}